\documentclass[twocolumn]{aastex61}

\received{}
\revised{}
\accepted{}
\submitjournal{ApJ}

\shorttitle{Carbon Chains in Low-mass Protostars}
\shortauthors{Law et al.}

\begin{document}

\title{Carbon Chain Molecules Toward Embedded Low-Mass Protostars\footnote{Based on observations carried out under project numbers 003-14 and 006-13 with the IRAM 30 m Telescope. IRAM is supported by INSU/CNRS (France), MPG (Germany), and IGN (Spain).}}

\correspondingauthor{Charles J. Law}
\email{charles.law@cfa.harvard.edu}

\author[0000-0002-0786-7307]{Charles J. Law}
\affiliation{Harvard-Smithsonian Center for Astrophysics, 60 Garden St., Cambridge, MA 02138, USA}

\author{Karin I. {\"O}berg}
\affiliation{Harvard-Smithsonian Center for Astrophysics, 60 Garden St., Cambridge, MA 02138, USA}

\author{Jennifer B. Bergner}
\affiliation{Harvard University Department of Chemistry and Chemical Biology, Cambridge, MA 02138, USA}

\author{Dawn Graninger}
\affiliation{Harvard-Smithsonian Center for Astrophysics, 60 Garden St., Cambridge, MA 02138, USA}

\begin{abstract}

Carbon chain molecules may be an important reservoir of reactive organics during star and planet formation. Carbon chains have been observed toward several low-mass young stellar objects (YSOs), but their typical abundances and chemical relationships in such sources are largely unconstrained. We present a carbon chain survey toward 16 deeply embedded (Class 0/I) low-mass protostars made with the IRAM 30 m telescope. Carbon chains are found to be common at this stage of protostellar evolution. We detect CCS, CCCS, HC$_3$N, HC$_5$N, l-C$_3$H, and C$_4$H toward 88\%, 38\%, 75\%, 31\%, 81\%, and 88\% of sources, respectively. Derived column densities for each molecule vary by one to two orders of magnitude across the sample. As derived from survival analysis, median column densities range between 1.2$\thinmuskip=-3mu\times 10^{11}$ cm$^{-2}$ (CCCS) and 1.5$\thinmuskip=-3mu\times 10^{13}$ cm$^{-2}$ (C$_4$H) and estimated fractional abundances with respect to hydrogen range between 2$\thinmuskip=-3mu\times 10^{-13}$ (CCCS) and 5$\thinmuskip=-3mu\times 10^{-11}$ (C$_4$H), which are low compared to cold cloud cores, warm carbon chain chemistry (WCCC) sources, and protostellar model predictions. We find significant correlations between molecules of the same carbon chain families, as well as between the cyanpolyynes (HC$_{\rm n}$N) and the pure hydrocarbon chains (C$_{\rm n}$H). This latter correlation is explained by a closely-related production chemistry of C$_{\rm{n}}$H and cyanpolyynes during low-mass star formation.

\end{abstract}

\keywords{astrobiology --- astrochemistry --- 
ISM: abundances --- ISM: molecules --- stars: protostars}



\section{Introduction} \label{sec:intro}

Carbon chain molecules constitute an important reservoir of reactive organic material in interstellar clouds \citep[e.g.,][]{Winnewisser79, Suzuki92, Ohishi98} and in at least some protostellar envelopes \citep{Sakai08, Sakai09, Sakai13}. These molecules may eventually be incorporated into protoplanetary disks \citep{Visser09, Visser11} and then later into planetesimals and planets, seeding incipient planets with complex organic material. Carbon chains are thus expected to be abundant throughout star formation and potentially during planet formation as well \citep{Qi13, Bergin16}. In this study, we present an inventory of such carbon chains toward low-mass protostars, the progenitors of the majority of planetary systems.

Carbon chain molecules were first observed in the dark cloud TMC-1 in the form of cyanopolyynes \citep{Little77, Broten78, Kroto78}. Since then, they have been discovered in many other molecular cloud lines of sight \citep{Suzuki92, Pratap97, Hirota04, Pardo05, Gupta09}. Their presence in such environments is well-explained by efficient low-temperature ion-molecule reactions in the gas phase \citep{Herbst89, Ohishi98}. This kind of chemistry is not expected to be very efficient once the cloud core collapses to form a protostar since by that time a large fraction of the carbon atoms have become fixed as CO. Moreover, models predict that many carbon chain molecules produced in the early, cold core stages are lost by reactions with He$^{+}$, H$^+$, and H$_3^+$ on timescales of $10^6 \,\rm{yr}$ \citep{Suzuki92}, as well as via faster depletion onto dust grains \citep{Hassel08}.

In light of the above observations and model predictions, the detections of abundant gaseous carbon chains (e.g.,~C$_{\rm{n}}$H, HC$_{\rm{n}}$N) in the lukewarm ($T \approx 30\,\rm{K}$) corino of low-mass protostar L1527 in 2008 by \citeauthor{Sakai08} came as a surprise. Detections toward a second source in the following year \citep{Sakai09} suggested that carbon chains may be common toward some classes of protostars. The abundances could not be accounted for by inheritance or cold, {\it in situ} ion chemistry and \citet{Sakai08} instead proposed that the observed carbon chains form in the lukewarm protostellar envelopes when methane, a common interstellar ice, sublimates at 25 K \citep{Oberg08}. This sublimation temperature is higher than that of CO (${\sim}20\,\rm{K}$) but is significantly lower than that of H$_2$O (${\sim}100\,\rm{K}$). Hence, CH$_4$ can be abundant in a warm region which is somewhat extended around the protostar \citep{Hassel08}. In the gas phase, CH$_4$ then reacts with C$^+$ to efficiently form carbon chains in the dense region heated by the emerging protostar (see \citet{Hassel08} for a detailed description of the intermediary reactions responsible for carbon chain production). The term ``warm carbon chain chemistry" (WCCC) was coined to describe this new type of source and chemistry. The WCCC hypothesis has since been validated by chemical models \citep{Hassel11, Aikawa12}. 

The occurrence rate of WCCC sources, i.e. protostellar envelopes with large abundances of carbon chains, is not well constrained. Four WCCC sources have been identified to date, and they include: L1527 \citep{Sakai08}, B228 \citep{Sakai09}, TMC-1A \citep{Aso15}, and L483 \citep{Hirota09ApJ...699..585H, Oya17}. \citet{Sakai08, Sakai09} suggest that these sources have a different chemical history compared to the typical protostar, which includes excess production of CH$_4$ in the protostellar stage. These WCCC sources have been well-studied and are well-explained by warm-up models that include CH$_4$ desorption \citep{Hassel08, Hassel11}. However, many other protostars are known to host moderate abundances of carbon chains in their envelopes \citep[e.g.,][]{Gregorio06, Graninger16, Lindberg16} and may present a more typical inventory of carbon chains during low-mass star formation. To assess the different possible organic inventories during planet formation, it is clearly important to understand both the extreme cases that the WCCCs present, and the distribution of carbon chains around more `typical' protostars. %

In this paper, we present the results of a carbon chain survey toward a sample of 16 embedded solar-type protostars using the IRAM 30 m telescope. In Section \ref{sec:observations}, we describe the observations and data reduction. In Section \ref{sec:results}, we present column densities, rotational temperatures, and estimated abundances for the observed carbon chains, as well as sample statistics and correlations. In Section \ref{sec:discussion}, we discuss molecular formation pathways and compare our results with cold clouds and carbon chain-rich protostars and then summarize our conclusions in Section \ref{sec:conclusions}.

\section{Observations} \label{sec:observations}

Detailed information about source selection and observational setup can be found in \citet{Graninger16}. In brief, 16 Class 0/I YSOs were chosen from the Spitzer \textit{c2d} ice sample of \citet{Boogert08}. All sources had an infrared spectral index, $\alpha_{\rm{IR}}$, defined as the slope between 2 and 24 $\mu$m, in excess of 0.3, which designates Class 0/I sources \citep{Wilking01}. Selections were based solely on their northern hemisphere locations and ice abundances (Table \ref{tab:source_info}). All sources were observed with the IRAM 30 m telescope using the EMIR 90 GHz receiver and the Fourier Transform Spectrometer (FTS) backend, which contains two 8~GHz sidebands with an 8~GHz gap in-between. Six of the sources (B1-a, B5 IRS1, L1489 IRS, IRAS 04108+2803, IRAS 03235+3004, and SVS 4-5) were observed on 2013 June 12--16 at 93--101~GHz and 109--117~GHz. The remaining sources were observed on 2014 July 23--28 at 92--100~GHz and 108--116~GHz. The spectral resolution for both sets of observations was 200 kHz (${\sim}0.55$ km s$^{-1}$) and the sideband rejection was $-15\rm{\,dB}$ \citep{Carter12}. The telescope beam size varies from $27^{\prime \prime}$ at 92~GHz to $21^{\prime \prime}$ at 117~GHz. As reported in \citet{Bergner17}, the rms values span 2--7~mK with the exception of the highest frequency spectral window.

\begin{deluxetable*}{lcccccccc}[!htp]
\tablecaption{Source Information of the 16-object \textit{c2d} Embedded Protostellar Sample with Ice Detections\label{tab:source_info}}
\tablehead{[-.3cm]
\colhead{Source} & \colhead{R.A.} & \colhead{Dec} & \colhead{Cloud} & \colhead{$\rm{L}_{\rm{bol}}$} & \colhead{$\rm{M}_{\rm{env}}$} & \colhead{$\alpha_{\rm{IR}}^a$} & \colhead{N(H$_2$O)$^a$} & \colhead{Systemic Velocity$^h$} \\[-.1cm]
& (J2000.0) & (J2000.0) & & $\left(L_{\odot}\right)$ & $\left(M_{\odot}\right)$ & &  $\left(10^{18} \, \rm{cm}^{-2}\right)$ & $\left(\rm{km}\,\rm{s}^{-1}\right)$}
\startdata
L1448 IRS1 & 03:25:09.44 & 30:46:21.7 & Perseus & 17.0$^c$ & 16.3$^c$ & 0.34 & 0.47 $\pm$ 0.16 & 3.4\\
IRAS 03235+3004$^b$ & 03:26:37.45 & 30:15:27.9 & Perseus&1.9$^c$ &2.4$^c$&1.44 & 14.48 $\pm$ 2.26 & 5.5\\
IRAS 03245+3002 & 03:27:39.03&30:12:59.3& Perseus & 7.0$^c$ & 5.3$^c$ & 2.70 & 39.31 $\pm$ 5.65 & 5.0\\
L1455 SMM1 & 03:27:43.25 & 30:12:28.8 & Perseus & 3.1$^c$ & 5.3$^c$ & 2.41& 18.21 $\pm$ 2.82 & 5.1\\
L1455 IRS3 & 03:28:00.41 & 30:08:01.2 & Perseus & 0.32$^c$ & 0.2$^d$ & 0.98 & 0.92 $\pm$ 0.37 & 5.0\\
IRAS 03254+3050 & 03:28:34.51 & 31:00:51.2 & Perseus & $-$ & 0.3$^c$ & 0.90 & 3.66 $\pm$ 0.47 & 7.2\\
IRAS 03271+3013 & 03:30:15.16 & 30:23:48.8 & Perseus & 0.8$^c$ & 1.2$^c$ & 2.06 & 7.69 $\pm$ 1.76 & 5.5\\
B1-a$^b$ & 03:33:16.67 & 31:07:55.1 & Perseus & 1.3$^c$ & 2.8$^c$ &1.87 & 10.39 $\pm$ 2.26 & 6.4 \\
B1-c & 03:33:17.89 & 31:09:31.0 & Perseus & 3.7$^c$ &17.7$^c$ &2.66 & 29.55 $\pm$ 5.65 & 6.5\\
B5 IRS 1$^b$  & 03:47:41.61 & 32:51:43.8&Perseus& 4.7$^c$ & 4.2$^c$ & 0.78 & 2.26 $\pm$ 0.28 & 10.1\\
L1489 IRS$^b$ & 04:04:43.07 & 26:18:56.4 & Taurus & 3.7$^e$ & 0.1$^f$  & 1.10 & 4.26 $\pm$ 0.51 & 7.7\\
IRAS 04108+2803$^b$ & 04:13:54.72 & 28:11:32.9 & Taurus & 0.62$^e$ & $-$ & 0.90 & 2.87 $\pm$ 0.40 & 7.2\\
HH 300 & 04:26:56.30 & 24:43:35.3 & Taurus & 1.27$^e$ & $-$ & 0.79 & 2.59 $\pm$ 0.25 & 7.0\\
SVS 4-5$^b$ & 18:29:57.59 & 01:13:00.6 & Serpens & 38$^g$ & $-$ & 1.26 & 5.65 $\pm$ 1.13 &  7.9 \\
L1014 IRS & 21:24:07.51 & 49:59:09.0 & L1014 & $-$ & $-$ & 1.28 & 7.16 $\pm$ 0.91 & 4.4\\
IRAS 23238+7401 & 23:25:46.65 & 74:17:37.2 & CB244 & $-$ & $-$ & 0.95 & 12.95 $\pm$ 2.26 & 4.5\\
\enddata
\tablecomments{Adapted from \citet{Graninger16, Bergner17}. \\ References: $^a$\citet{Boogert08}, $^b$Sources were observed by \citet{Oberg14}, $^c$\citet{Hatchell07}, $^d$\citet{Enoch09}, $^e$\citet{Furlan08}, $^f$\citet{Brinch07}, $^g$\citet{Pontoppidan04}, $^h$this study.}
\end{deluxetable*}

CLASS\footnote{\url{http://www.iram.fr/IRAMFR/GILDAS}} was used to reduce the spectra. Global baselines were fit to each 4~GHz spectral chunk using four to seven line-free windows. Individual scans were baseline subtracted and averaged. Ruze's equation was used to modify the beam efficiency with a scaling factor of 0.861 and sigma of 63.6 microns, which led to beam efficiencies at the first, middle, and last channel of 0.81, 0.80, and 0.78, respectively. The scaling factor and sigma value are the same as used in \citet{Bergner17}. With a forward efficiency of 0.95, antenna temperature was then converted to main beam temperature, $T_{\rm{mb}}$. Literature source velocities were initially used to convert spectra to an approximate rest frequency with fine-tuning adjustments made using the CS $J=2-1$ transition and are listed in Table \ref{tab:source_info}.

\section{Results} \label{sec:results}

\subsection{Molecule Detections}

\begin{figure*}[htp!]
\centering
\includegraphics[scale=0.7]{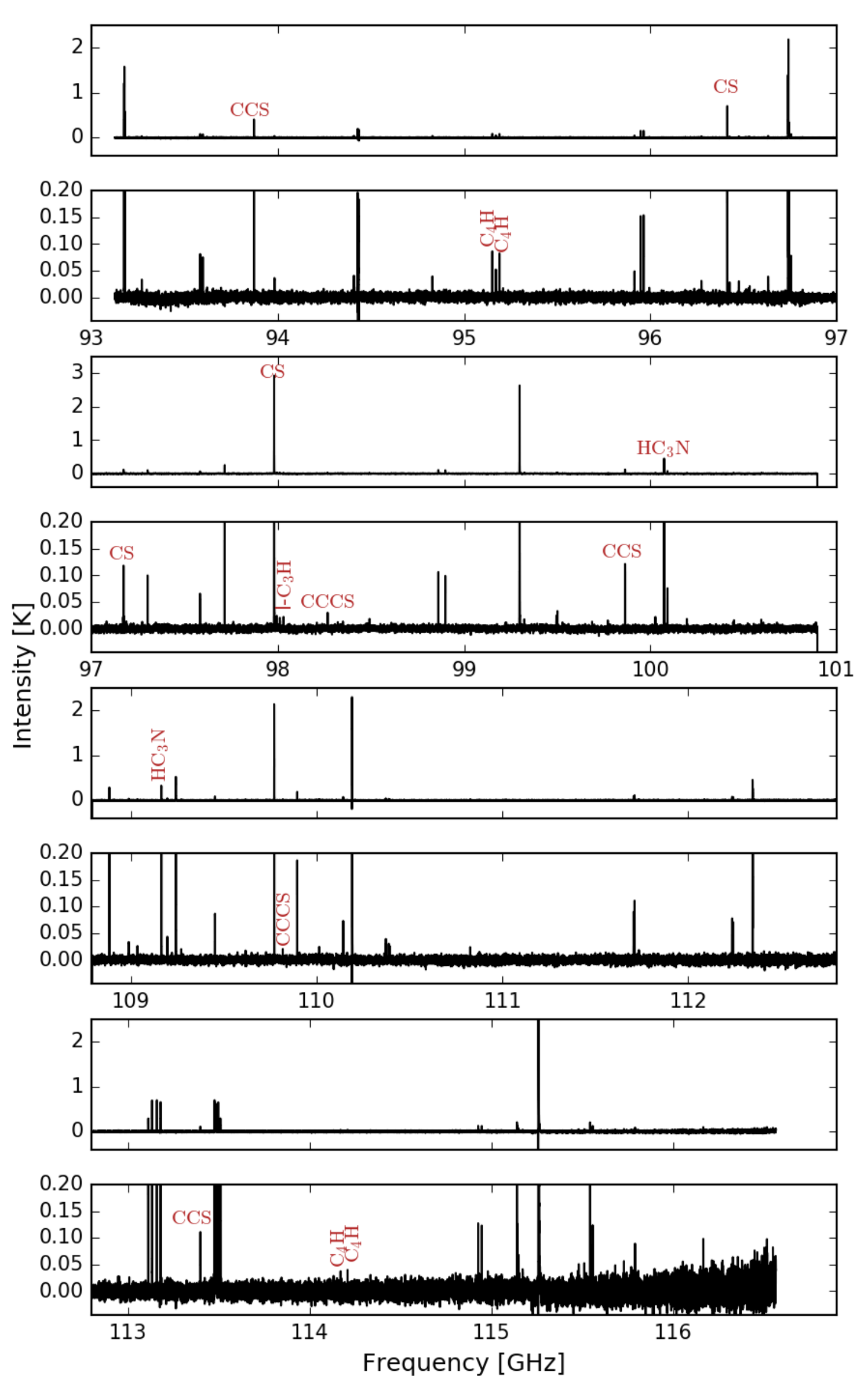}
\caption{Spectra at 93--101 and 109--117~GHz toward B1-a with key carbon chain lines highlighted. Most of the remaining lines have been previously identified to methanol, complex organic molecules (COMs), CO isotopologues, and CN and are the subject of previous studies \citep[cf.][]{Oberg14, Graninger16, Bergner17}. \label{fig:chains_labeled}}
\end{figure*}

The complete spectrum toward one of our sources, B1-a, with carbon chain identifications is shown in Figure \ref{fig:chains_labeled}.
The spectrum contains a number of carbon chain molecules that have been detected toward other high- or low-mass protostars. We focus on six carbon chains, which are commonly seen in our low-mass protostellar survey: sulfur-bearing chains CCS and CCCS, cyanopolyynes HC$_3$N and HC$_5$N, and pure hydrocarbons l-C$_3$H and C$_4$H. We also include CS and its isotopologues in our analysis to investigate the sulfur carbon chain chemistry. The frequencies, line strengths, and upper-state energies of the observed lines are summarized in Table \ref{tab:line_info_methods}. 
Line characteristics come from the JPL\footnote{\url{http://spec.jpl.nasa.gov}} and CDMS\footnote{\url{http://www.astro.uni-koeln.de/cdms/catalog}} catalogs. We note that catalogs sometime differ in their consideration of nuclear spin in the $S$ values and partition functions. In our case this affects the $J=2-1$ transition of C$^{33}$S, whose CDMS $S$ value is fourfold that found in some other catalogs, e.g. the SLAIM catalog. Care therefore has to be taken to use matching $S$ values and partition functions; in this particular case, we took both from the CDMS catalog.
A molecule is considered to be detected if: (1) we observed at least one line with a $5\sigma$ detection or two lines with $3\sigma$ detections; (2) it is free of confusion from other common interstellar or YSO molecular lines; and (3) upper limits of non-detected lines are consistent with predicted populations from detected lines. The treatment of upper limits is discussed in further detail in Section \ref{sec:rot_diag_and_temp}. In our survey, overlapping lines were typically not a problem, since even the most line-rich sources in our sample are line-poor relative to hot cores. Specifically, our most line-dense sources, such as B1-a, have typical spectral line densities of ${\sim}$6--8 lines per GHz, while hot cores can present 10s or even 100s of lines per GHz and factors of a few greater line widths \citep{Tercero10, Jorgensen16, Bonfand17}. Thus, provided that there are no competing line identifications, a single line is sufficient to claim a detection. All detected lines have upper excitation energies $<150\,\rm{K}$, consistent with the expected $<100$~K temperatures of low-mass protostars.

\begin{deluxetable*}{lcccc}[!htp]
\tablecaption{Carbon Chain Molecular Lines Observed Toward Our Protostellar Sample\label{tab:line_info_methods}}
\tablehead{[-.3cm]
\colhead{Molecule} & \colhead{Transition} & \colhead{$\nu$ (MHz)} & \colhead{$S$\tablenotemark{$a$}} & \colhead{$E_u$ (K)} \\[-.55cm]}
\startdata
C$^{34}$S .\,.\,.\,.\,.\,.\,.\,.\,.\,.\,.\,. & $J = 2-1$ &  96412.950 & 2.00 & 6.25 \\
C$^{33}$S .\,.\,.\,.\,.\,.\,.\,.\,.\,.\,.\,. & $J = 2-1$ &  97172.064 & 7.98 & 6.31 \\
CS .\,.\,.\,.\,.\,.\,.\,.\,.\,.\,.\,.\,.\,.  & $J = 2-1$ &  97980.953 & 2.00 & 7.05 \\
CCS \ldots \ldots \ldots \ldots & $J_N = 8_7 - 7_6$ & 93870.107 & 7.97 & 19.89 \\
 & $J_N = 7_8 - 6_7$ & 99866.521 & 6.87 &  	28.14 \\
 & $J_N = 8_9 - 7_8$ & 113410.186 & 7.89 & 33.58 \\
CCCS .\,.\,.\,.\,.\,.\,.\,.\,.\,. & $J = 17-16$ & 98268.518 & 17.00 & 42.45 \\
 & $J = 19-18$ & 109828.293 & 19.00 & 52.71 \\
HC$_3$N .\,.\,.\,.\,.\,.\,.\,.\,.\,. & $J = 11 -10$ & 100076.392 & 11.00 & 28.82 \\
 & $J=12-11$ & 109173.634 & 12.00 & 34.06 \\
HC$_5$N .\,.\,.\,.\,.\,.\,.\,.\,.\,. & $J=35-34$ & 93188.471& 34.99 & 80.50 \\
 & $J=36-35$ & 95850.714 & 35.99 & 85.10 \\
 & $J=37-36$ & 98512.933 & 36.99 & 89.83 \\
 & $J=41-40$ & 109161.555 & 40.99 & 110.02 \\
 & $J=42-41$ & 111823.643 & 41.99 & 115.39 \\
 & $J=43-42$ & 114485.704 & 42.99 & 120.88 \\
l-C$_3$H \ldots \ldots \ldots \ldots & $J_{F,\,l=e} = \left(\frac{9}{2}\right)_5 - \left(\frac{7}{2}\right)_4$ & 97995.166 & 4.89 & 12.54 \\
 & $J_{F,\,l=e} = \left(\frac{9}{2}\right)_4 - \left(\frac{7}{2}\right)_3$ & 97995.913 & 3.89 & 12.54 \\
 & $J_{F,\,l=f} = \left(\frac{9}{2}\right)_5 - \left(\frac{7}{2}\right)_4$ & 98011.611 & 4.87 & 12.54 \\
 & $J_{F,\,l=f} = \left(\frac{9}{2}\right)_4 - \left(\frac{7}{2}\right)_3$ & 98012.524 & 3.87 & 12.54 \\
C$_4$H\tablenotemark{$b$}  \ldots \ldots \ldots \ldots & $N_F = 10_{11} - 9_{10}$ & 95150.397 & 10.23 & 25.11 \\
 & $N_F = 10_{9} - 9_{8}$ & 95188.946 & 8.36 & 25.13 \\
 & $N_F = 12_{12} - 11_{11}$ & 114182.512 & 11.17 & 35.62 \\
 & $N_F = 12_{11} - 11_{10}$ & 114221.040 & 10.23 & 35.64 \\
\enddata
\tablenotetext{a}{Line strength.\\[-.4cm]}
\tablenotetext{b}{C$_4$H transitions were reported in \citet{Graninger16}.\\[-.4cm]}
\tablecomments{Rest frequency, line strength, and upper-state energy are taken from the JPL and CDMS catalogs.}
\end{deluxetable*}

We detect CS, CCS, CCCS, HC$_3$N, HC$_5$N, l-C$_3$H, and C$_4$H in 16, 14, 6, 12, 5, 13, and 14 sources, respectively, which corresponds to detection percentages of 100\%, 88\%, 38\%, 75\%, 31\%, 81\%, and 88\%. We also report additional tentative detections of CCS and CCCS each in one source and HC$_5$N in two sources, based on single $3$--$4\sigma$ line detections.

\begin{figure*}[!htp]
\centering
\includegraphics[scale=0.6]{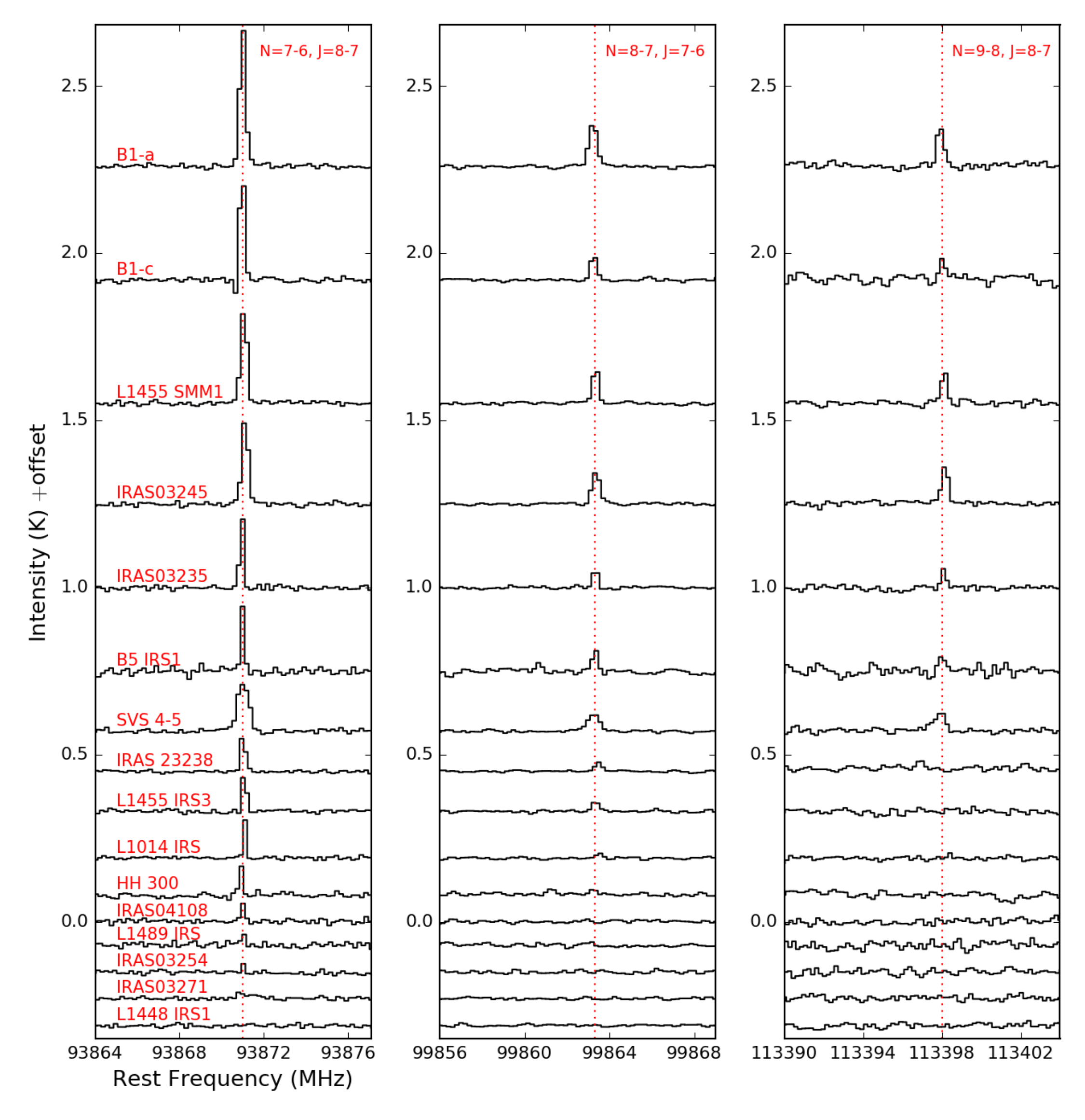}
\caption{Zoomed-in spectra of CCS toward the low-mass YSO sample. Rest frequencies derived assuming the characteristic velocity of each source. Figures showing spectra for the remaining carbon chains are available in the figure set. (\textit{The complete figure set (7 images) containing spectra of all species and lines is available in the online journal}.)}
\label{fig:CCS_detections_all_sources}
\end{figure*}

In Figure \ref{fig:CCS_detections_all_sources}, we zoom in on the frequency ranges containing CCS lines to visualize line shapes in different sources and the diversity of line strengths across the sample. Spectral windows containing lines of the remaining molecules are presented in Figure Set 2, which is available in the electronic edition of the journal. Since we have a velocity resolution of ${\sim}0.55$ km s$^{-1}$ and a typical line width of ${\sim}1.15$~km~s$^{-1}$, many lines are barely resolved, but most lines are consistent with Gaussian shapes. A few sources display small wings, most notably SVS 4-5, which may be due to low-velocity molecular outflows. We see no evidence for jets in any of the sources. Wings were visually identified and then manually excluded by restricting the Gaussian fitting range and hence are not included when determining the molecular abundances below.


\figsetgrpstart
\figsetgrpnum{2.1}
\figsetgrptitle{CCS Spectra}
\figsetplot{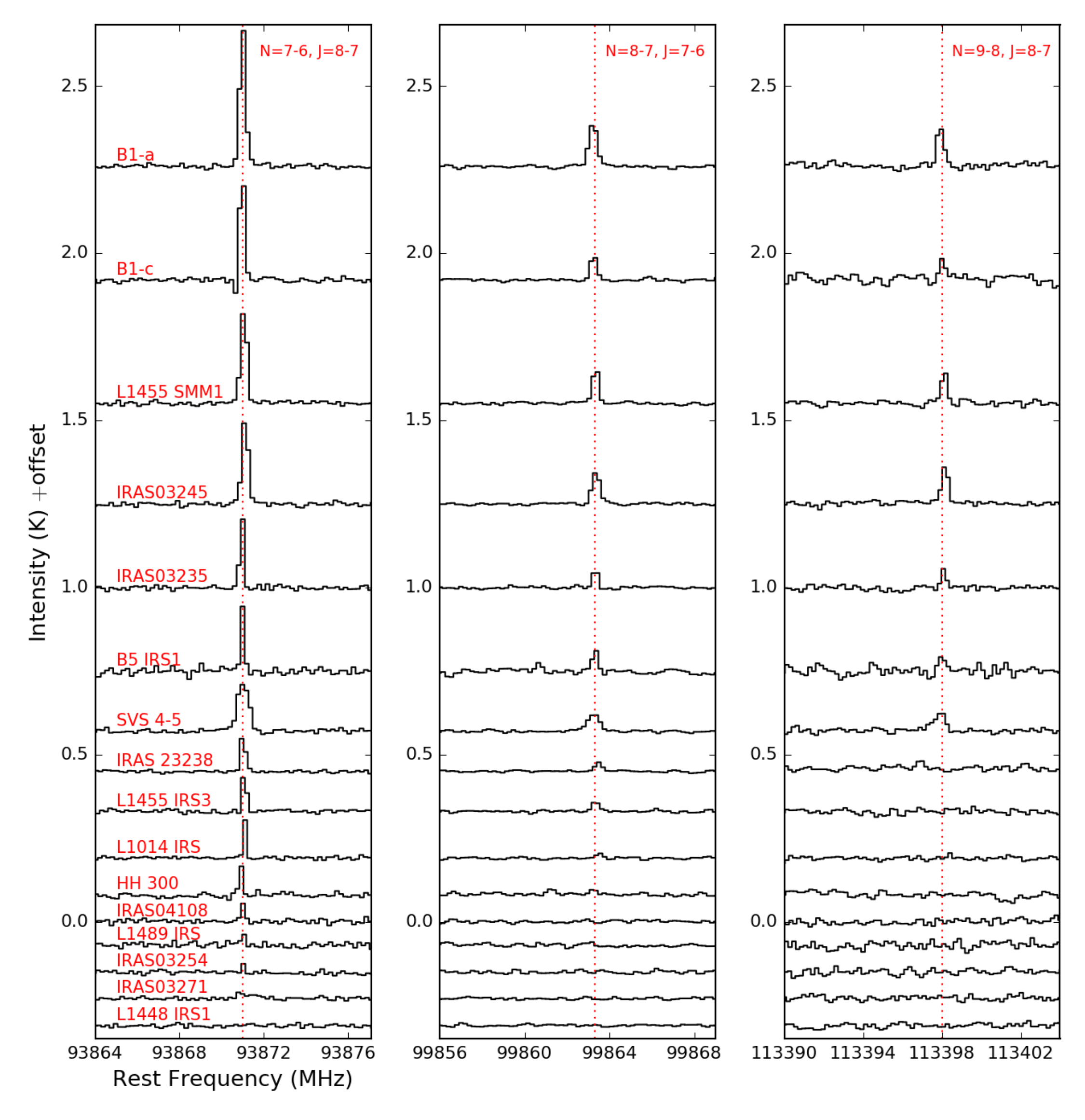}
\figsetgrpnote{Zoomed-in spectra of CCS toward the low-mass YSO sample. Rest frequencies derived assuming the characteristic velocity of each source.}
\figsetgrpend

\figsetgrpstart
\figsetgrpnum{2.2}
\figsetgrptitle{CS Spectra}
\figsetplot{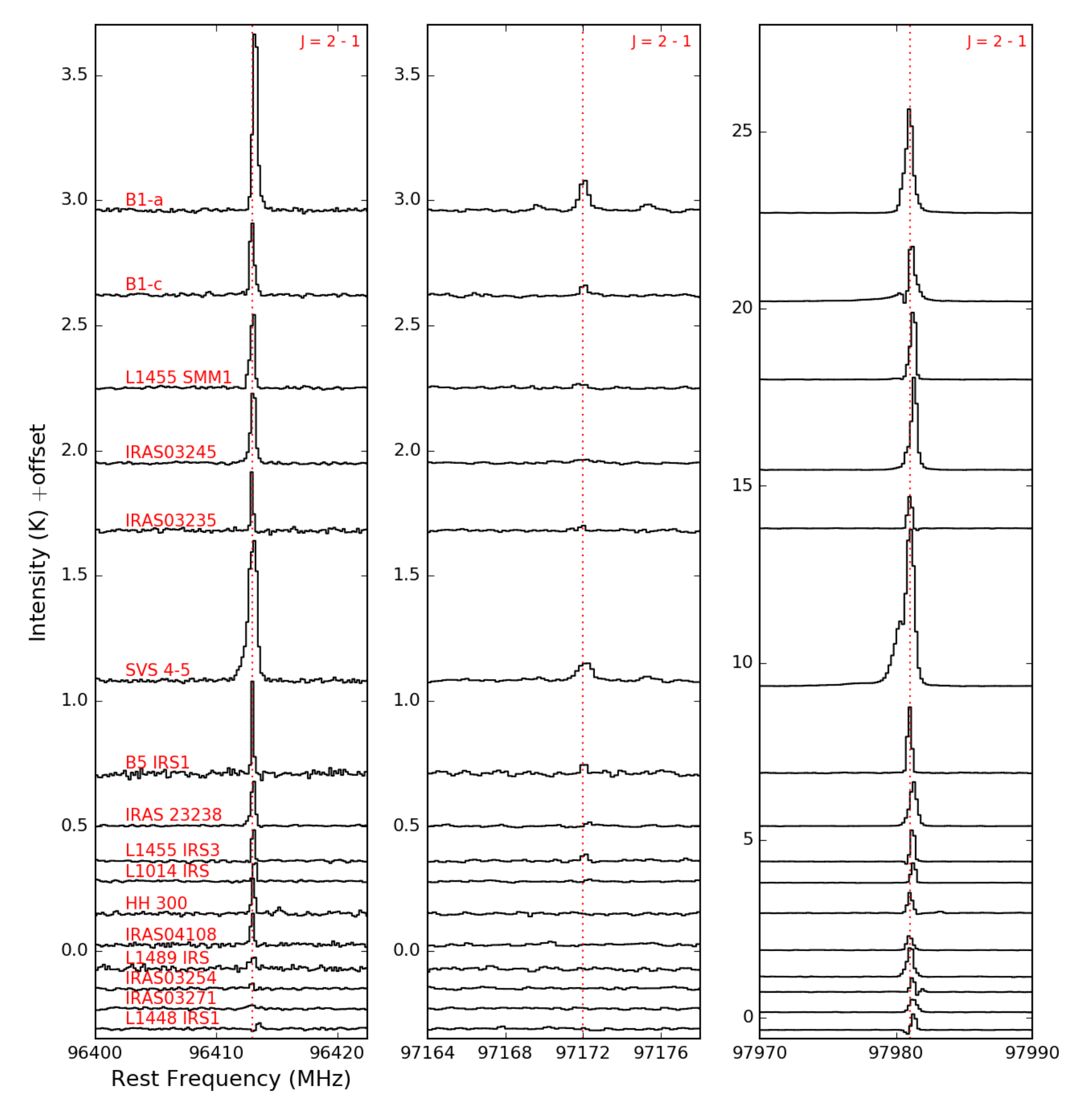}
\figsetgrpnote{Zoomed-in spectra of CS toward the low-mass YSO sample. Rest frequencies derived assuming the characteristic velocity of each source.}
\figsetgrpend

\figsetgrpstart
\figsetgrpnum{2.3}
\figsetgrptitle{CCCS Spectra}
\figsetplot{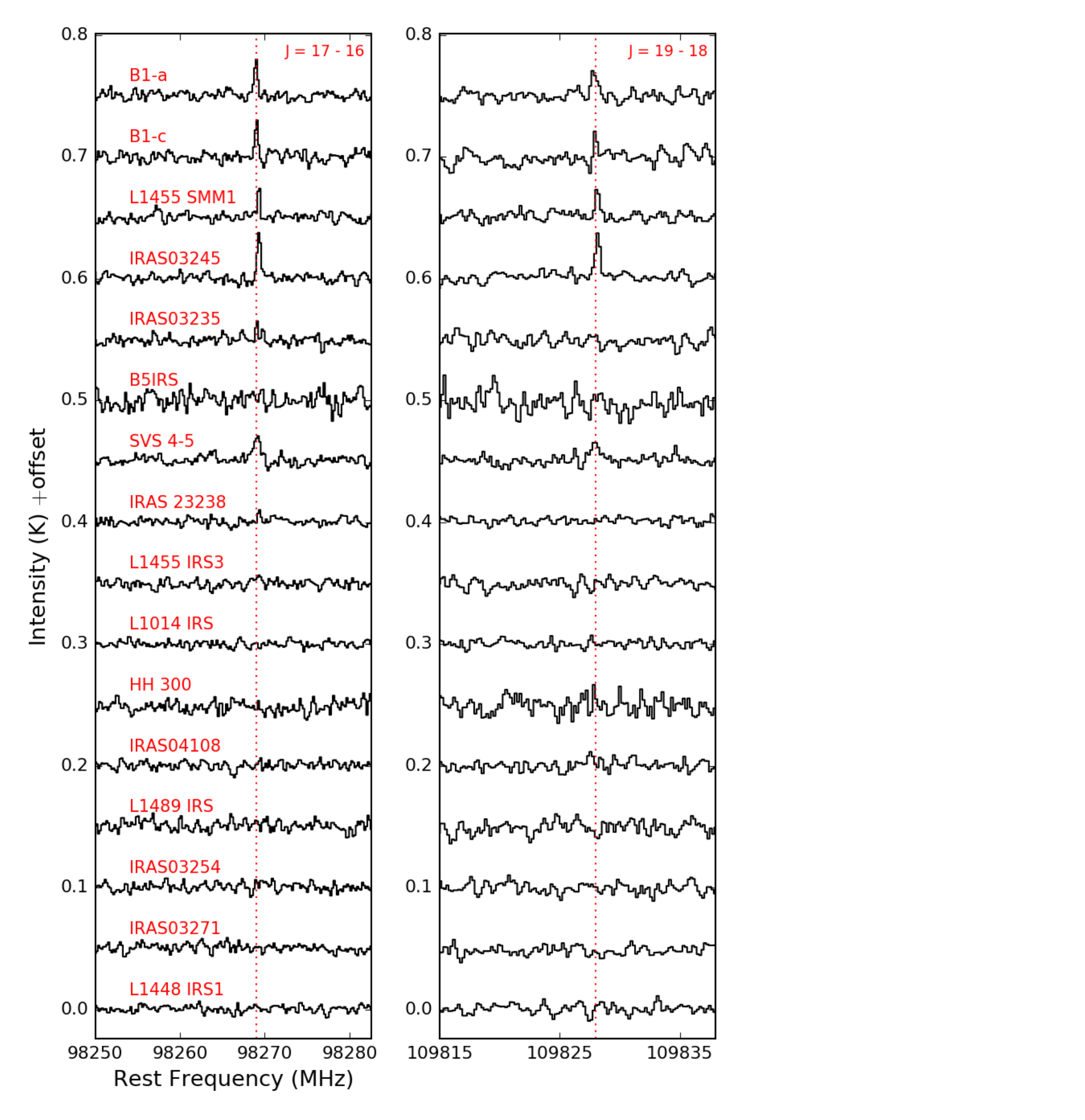}
\figsetgrpnote{Zoomed-in spectra of CCCS toward the low-mass YSO sample. Rest frequencies derived assuming the characteristic velocity of each source.}
\figsetgrpend

\figsetgrpstart
\figsetgrpnum{2.4}
\figsetgrptitle{C$_3$H Spectra}
\figsetplot{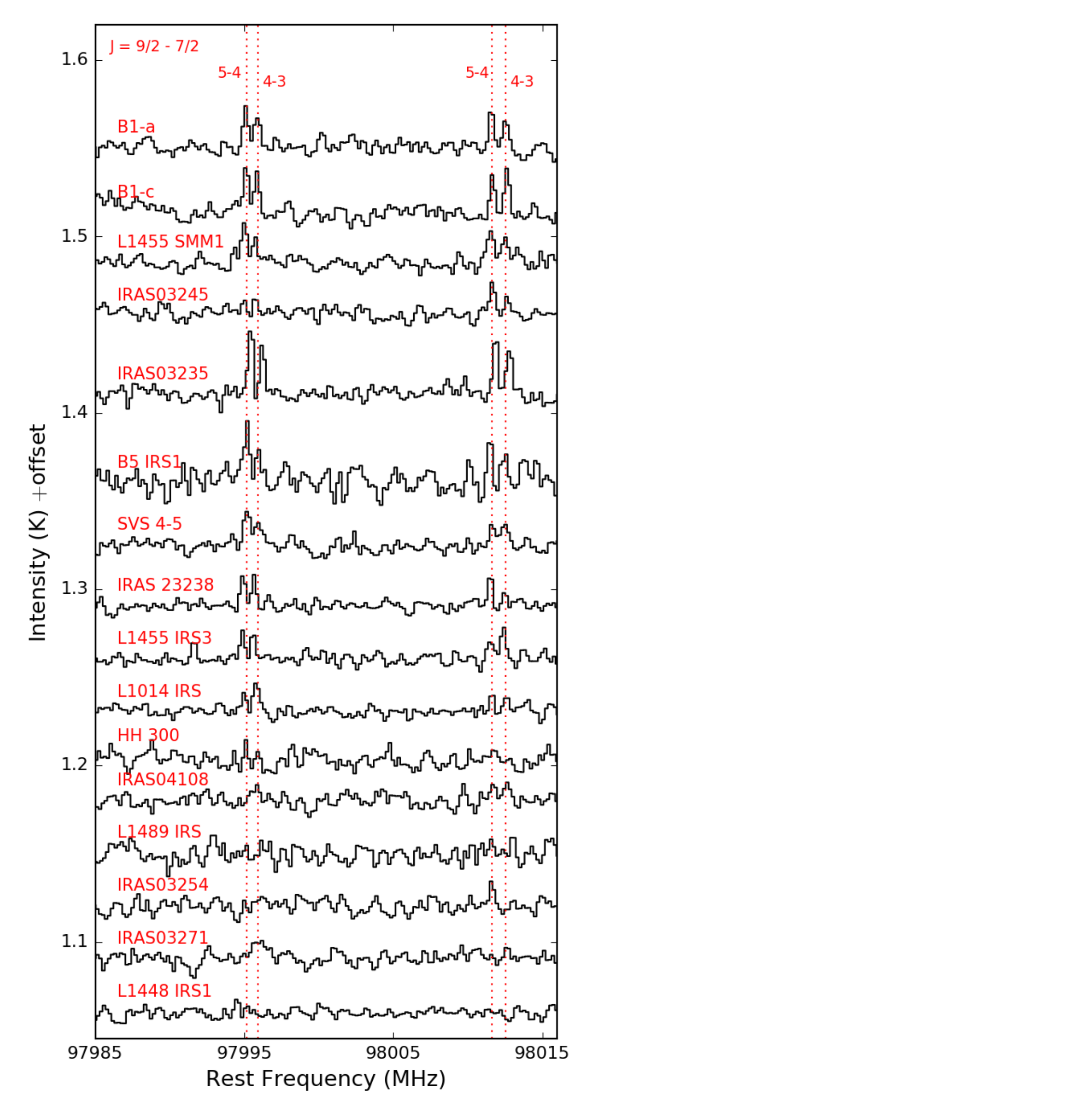}
\figsetgrpnote{Zoomed-in spectra of C$_3$H toward the low-mass YSO sample. Rest frequencies derived assuming the characteristic velocity of each source.}
\figsetgrpend

\figsetgrpstart
\figsetgrpnum{2.5}
\figsetgrptitle{HC$_3$N Spectra}
\figsetplot{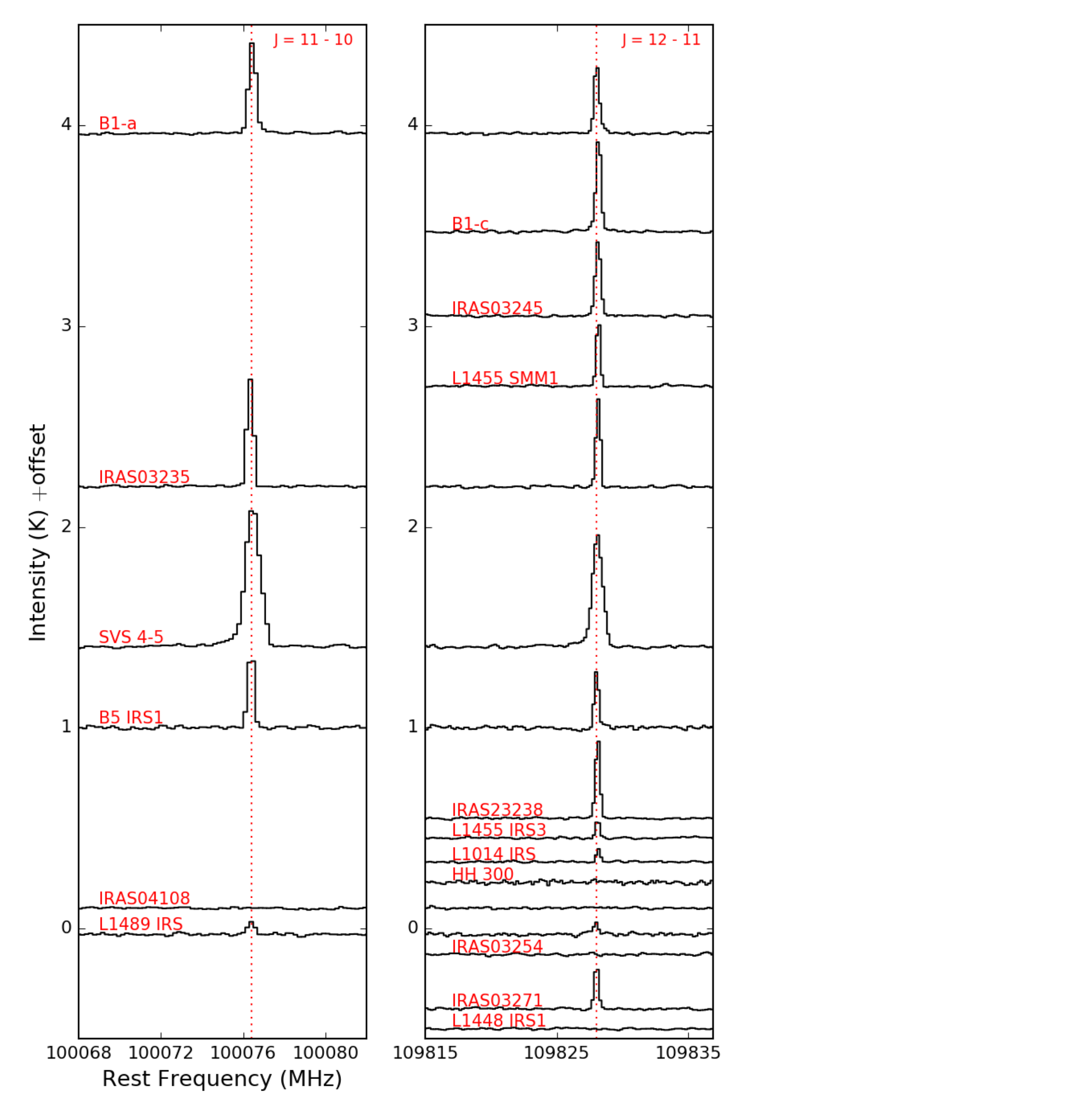}
\figsetgrpnote{Zoomed-in spectra of HC$_3$N toward the low-mass YSO sample. Rest frequencies derived assuming the characteristic velocity of each source.}
\figsetgrpend

\figsetgrpstart
\figsetgrpnum{2.6}
\figsetgrptitle{HC$_5$N Spectra, Part 1}
\figsetplot{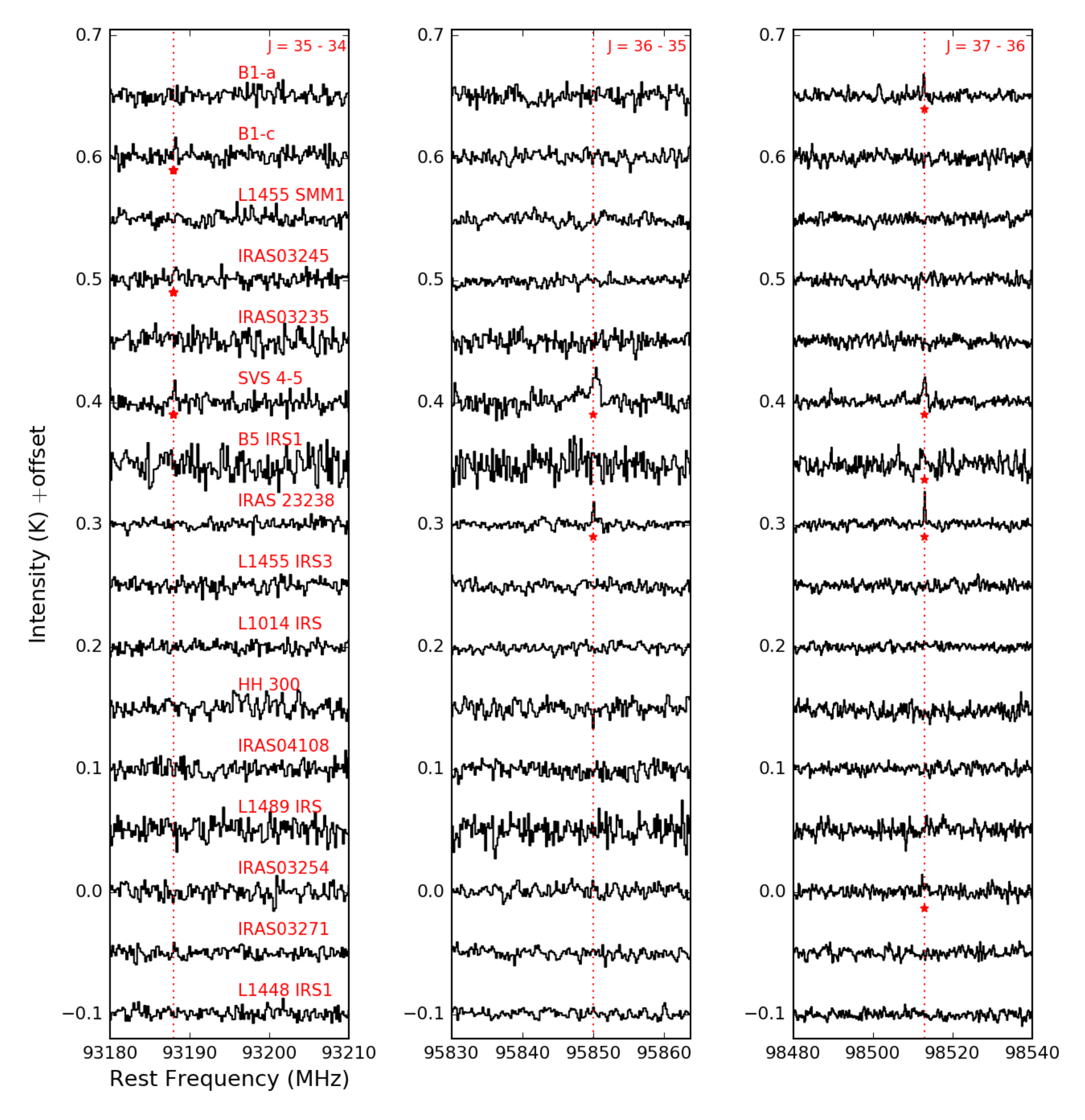}
\figsetgrpnote{Zoomed-in spectra of HC$_5$N toward the low-mass YSO sample. Rest frequencies derived assuming the characteristic velocity of each source. Individual line detections are highlighted with red stars.}
\figsetgrpend

\figsetgrpstart
\figsetgrpnum{2.7}
\figsetgrptitle{HC$_5$N Spectra, Part 2}
\figsetplot{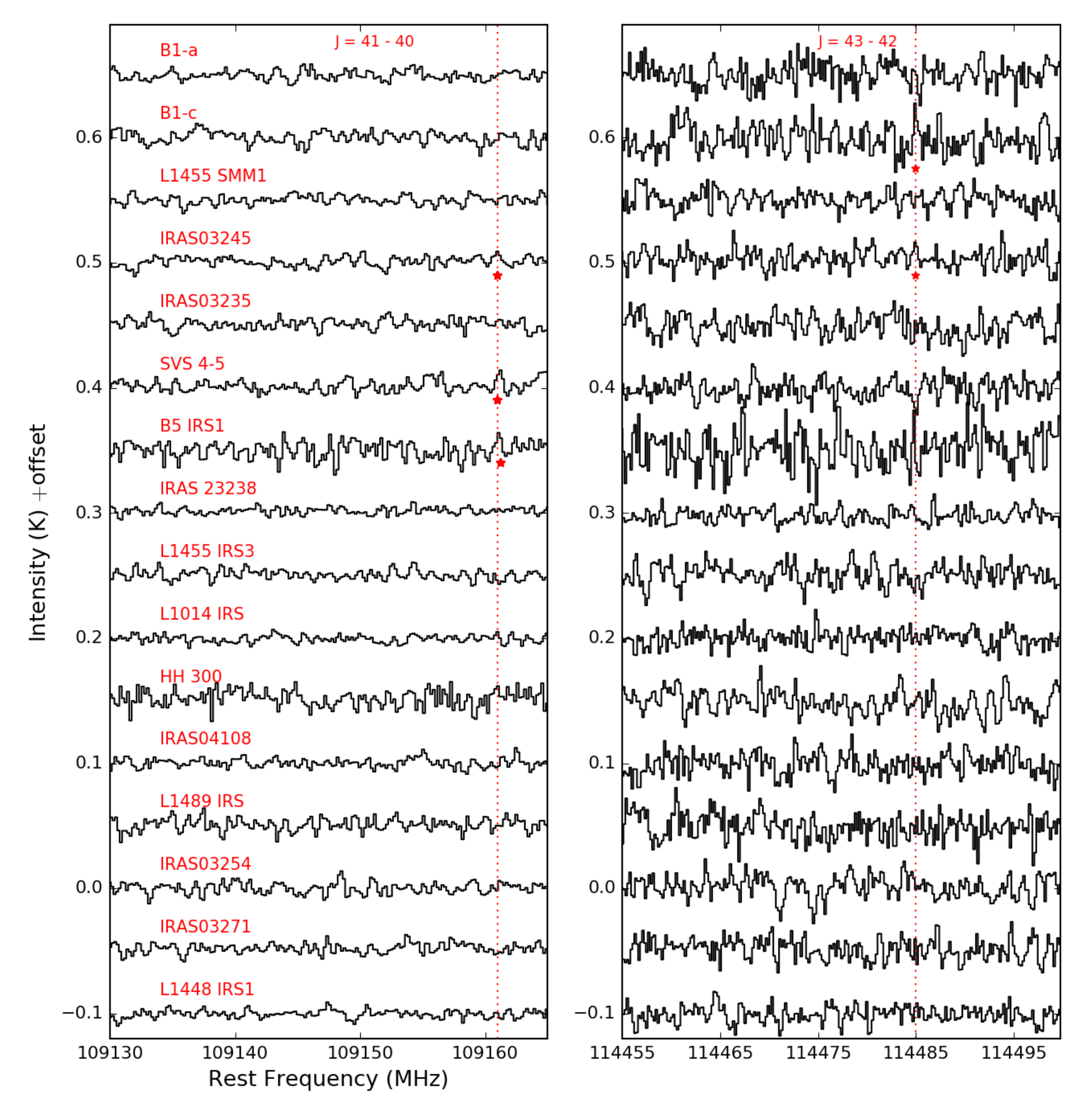}
\figsetgrpnote{(continued) Zoomed-in spectra of HC$_5$N toward the low-mass YSO sample. Rest frequencies derived assuming the characteristic velocity of each source. Individual line detections are highlighted with red stars.}
\figsetgrpend

\figsetend

\subsection{Rotational Diagrams and Temperatures}
\label{sec:rot_diag_and_temp}

Integrated line intensities were determined by fitting single Gaussian profiles to each spectral feature (excluding line wings) and are listed in Tables \ref{tab:CS_intensities}--\ref{tab:HC5N_intensities} in the Appendix. Features were fit assuming a zero-order polynomial baseline with a fixed central line frequency. We treated unresolved multiplets arising from the same species as a single line by combining degeneracies and intrinsic line intensities in the analysis below. In addition to the uncertainty in the line fit, we adopt a 10\% calibration uncertainty.

In cases of non-detections, $3\sigma$ line intensity upper limits were calculated as:

\begin{equation} \label{eq:4}
\sigma = \rm{rms} \times \frac{\rm{FWHM}}{\sqrt{n_{\rm{ch}}}}.
\end{equation}

\noindent In this case, we take the rms from a 40 km s$^{-1}$ spectral window containing the transition. The FWHM was taken to be the same as that of other lines of the same molecule, if such lines exist, or of closely related molecules in the same source. Here, the number of channels, $n_{\rm{ch}}$, across the FWHM was ${\sim}$FWHM/0.6 km s$^{-1}$ channel$^{-1}$. The FWHMs for all detected lines ($\geq 3\sigma$) are listed in Table \ref{tab:FWHMs_Detected_Lines_1} in the Appendix.
For molecules with multiple line detections spread over a range of energies, we used rotational diagrams to calculate rotational temperatures and column densities \citep{Goldsmith99}. Upper level populations for each line are given by:

\begin{equation} \label{eq:2}
\frac{N_u}{g_u} = \frac{3k_B \int T_{\rm{mb}} dV}{8 \pi^3 \nu \mu^2 S}
\end{equation}

\noindent where $g_u$ is the statistical weight of level $u$, $k_B$ is the Boltzmann constant, $\nu$ is the transition frequency, $\mu$ is the permanent dipole moment, $S$ is the intrinsic line strength, and $N_u$ is the column density in the upper level $u$. If we assume optically thin lines and local thermodynamic equilibrium (LTE), each molecule's total column density, $N_{\rm{tot}}$, and rotational temperature, $T_{\rm{rot}}$, in each source can be calculated from:

\begin{equation} \label{eq:3}
\frac{N_u}{g_u} = \frac{N_{\rm{tot}}}{Q(T_{\rm{rot}})} \exp{\left( -\frac{E_u}{T_{\rm{rot}}} \right)}
\end{equation}

\noindent where $Q(T_{\rm{rot}})$ is the rotational partition function and $E_u$ is the energy of the upper level $u$ \citep{Goldsmith99}. We also assume that all molecules contained within the beam can be characterized by a single temperature \citep[e.g.,][]{Gregorio06, Bisschop07b, Sakai08, Sakai09, Oberg14, Fayolle15, Bergner17}. 

We do not include effects related to absorption against the CMB or optical depths \citep[e.g.,][]{Suzuki92, Takano98}. At the frequencies and rotational temperatures we are considering, we find that the inclusion of the CMB contribution only leads to a $<10\%$ difference in the $J(T)$ term, as defined in Equation 3 in \citet{Takano98}, which is small compared to other sources of error. The optical depths for all species (expect CS), based our derived column densities and line widths, are small with $\tau \ll 1$.

Rotational diagrams for CCS are shown in Figure \ref{fig:Sample_rot_diag_CCS} and diagrams for all detected molecules are presented in the Appendix (Figures \ref{fig:appendix_CCCS_RD}--\ref{fig:appendix_HC5N_RD}). We determine rotational excitation temperatures for all sources and species with at least two line detections for a given molecule. The median rotational temperatures for CCS, CCCS, HC$_3$N, and C$_4$H are 9~K, 16~K, 13~K, and 10~K, respectively. These molecules all have relatively tight distributions of rotational temperatures, as indicated by their narrow quartile ranges. No rotational diagrams could be constructed for C$^{34}$S and l-C$_3$H. For HC$_5$N, only two sources had a sufficient number of detections to calculate rotational temperatures and of these only one source (SVS 4-5) had strong enough lines to obtain a well-constrained rotational temperature of 42~K. 

\begin{figure*}[!htp]
\centering
\includegraphics[scale=0.7]{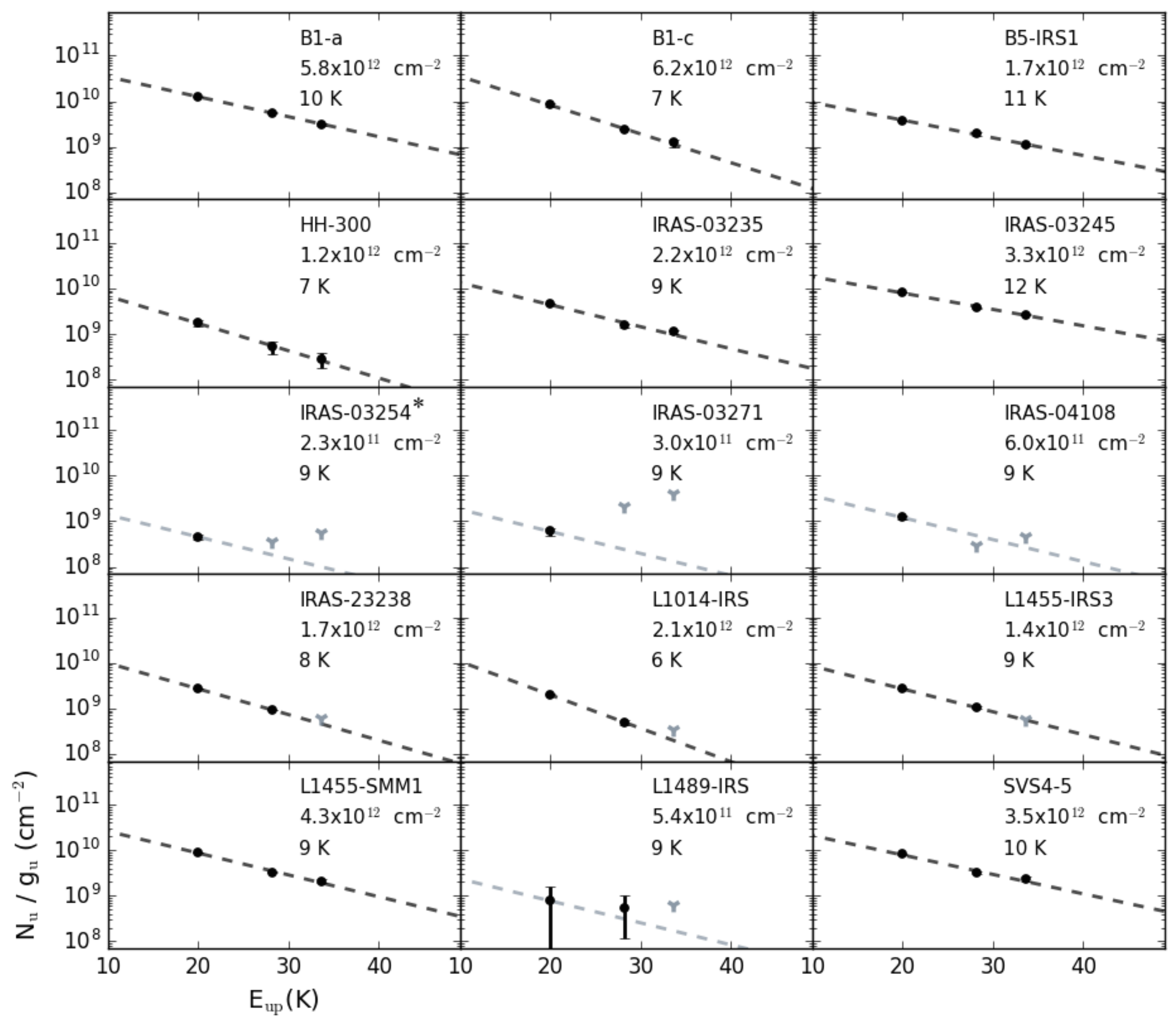}
\caption{Rotational diagrams for CCS. Black circles represent detections and gray triangles indicate upper limits. Black dashed lines are fits to the data. When a line was unable to be fit, a rotational temperature was assumed as described in the text and is shown as a gray dashed line. Error bars are smaller than the symbol for most cases. A tentative CCS detection in IRAS 03254 is denoted by an asterisk ($^*$).}
\label{fig:Sample_rot_diag_CCS}
\end{figure*}

\begin{deluxetable*}{lccccc}[!htp]
\tablecaption{Carbon Chain Rotational Temperatures in K for the 16 Sources\label{tab:Rotation_Temperatures}}
\tablehead{[-.3cm]
\colhead{Source} & \colhead{CCS} & \colhead{CCCS} & \colhead{HC$_3$N} & \colhead{HC$_5$N} & \colhead{C$_4$H\tablenotemark{$a$}}\\[-.55cm]} 
\startdata
L1448 IRS1 & $\textit{9 [2]}$ & $\textit{16 [6]}$ & $\textit{13 [1]}$ & $\textit{28 [6]}$ & 10 [2] \\
IRAS 03235+3004 & 9 [1] & $\textit{16 [6]}$  & 14 [3] & $\textit{28 [6]}$ & 10 [3] \\
IRAS 03245+3002 & 12 [2] & 18 [4] & $\textit{13 [1]}$ & $\textit{28 [6]}$ & 10 [4] \\
L1455 SMM1 & 9 [1] & \,\,\,\,\, 37 [11]$^*$ &  $\textit{13 [1]}$ & $\textit{28 [6]}$ & 11 [7] \\
L1455 IRS3 & 9 [1] & $\textit{16 [6]}$ & $\textit{13 [1]}$  & $\textit{28 [6]}$ & 10 [2] \\
IRAS 03254+3050 & $\textit{9 [2]}$ & $\textit{16 [6]}$ & $\textit{13 [1]}$ & $\textit{28 [6]}$ & \, $\textit{10 [1.6]}$  \\
IRAS 03271+3013 & $\textit{9 [2]}$  & $\textit{16 [6]}$ & $\textit{13 [1]}$ & $\textit{28 [6]}$ & 10 [3]  \\
B1-a & 10 [1] & 16 [6] & 12 [2] & $\textit{28 [6]}$ & 9.6 [1.7]  \\
B1-c & 7 [1] & 10 [6] & $\textit{13 [1]}$ & $\textit{28 [6]}$ & 10 [2]  \\
B5 IRS 1 & 11 [1] & $\textit{16 [6]}$ & 11 [2] & $\textit{28 [6]}$ & 9.5 [1.0]  \\
L1489 IRS & $\textit{9 [2]}$  & $\textit{16 [6]}$  & $\textit{13 [1]}$ & $\textit{28 [6]}$ & 10 [2]  \\
IRAS 04108+2803 & $\textit{9 [2]}$ & $\textit{16 [6]}$ & $\textit{13 [1]}$ &$\textit{28 [6]}$& 10 [2] \\
HH 300 & 7 [1]  & $\textit{16 [6]}$ & $\textit{13 [1]}$ & $\textit{28 [6]}$ & 10 [2]  \\
SVS 4-5 & 10 [1] & 12 [5] & 14 [3] & 42 [9]$^*$ & 15 [3] \\
L1014 IRS & 6 [1] & $\textit{16 [6]}$ & $\textit{13 [1]}$ & $\textit{28 [6]}$ & 6.9 [1.0]  \\
IRAS 23238+7401 & 8 [1] & $\textit{16 [6]}$ & $\textit{13 [1]}$ & $\textit{28 [6]}$ & 10 [1] \\ \hline
Species Median & $9.0^{10.0}_{7.5}$ & $16.0_{12.0}^{18.0}$ & $13.0_{11.8}^{14.0}$ & -- & $10.0_{10.0}^{10.0}$   \\ 
\enddata
\tablenotetext{$a$}{C$_4$H data are taken from \citet{Graninger16} with new C$_4$H detection from source IRAS 03254$+$3050.\\[-.4cm]}
\tablecomments{Uncertainties at the $1\sigma$ level are reported in brackets. $T_{\rm{rot}}$ values in italics are assumed rotational temperatures and are based on median sample temperature. Asterisks ($^*$) indicate unexpectedly high temperatures for CCCS and HC$_5$N in L1455 SMM1 and SVS 4-5, respectively; additional data are needed to confirm these temperatures.}
\end{deluxetable*}

Based on the results summarized in Table \ref{tab:Rotation_Temperatures}, there is a spread in species median temperatures across molecules of 9--16 K. There does not appear to be any significant trend between the rotational temperature of the sources and their line-richness. Specifically, the line-rich sources B1-a and IRAS 23238$+$7401, moderately line-dense source B5 IRS1, and line-poor source HH~300 all have approximately the same rotational temperature. For the sulfur-bearing chains, temperature increases with molecular size; the median CCCS rotational temperature is 80\% higher than the CCS median. This trend may be explained by either a chemical effect, i.e. over-production of longer sulfur-bearing carbon chains at higher temperatures, or more likely, an excitation effect, as is seen in dark clouds where longer carbon chains are less efficiently radiatively cooled \citep[e.g.,][]{Bell98}.


\subsection{Column Densities}

We calculate column densities without any assumptions about beam dilution, since carbon chains can be produced at a range of temperatures and the emission may therefore fill the telescope beam (see Section \ref{sec:rot_diag_and_temp} for further discussion). All column densities are reported in Table \ref{tab:column_density_large_table}. For species and sources with well-constrained rotational temperatures, we use Equations \ref{eq:2} and \ref{eq:3} to calculate column density. In calculating column densities for CCS, CCCS, HC$_3$N, and C$_4$H in sources with missing rotational temperatures, we adopt the sample median rotational temperature. As noted above, the narrow temperature quartile range for each species implies that using the sample median should not introduce any major errors. For l-C$_3$H, the column densities were calculated using the median C$_4$H rotational temperature reported by \citet{Graninger16} and for C$^{34}$S, the median CCS rotational temperature was used. These choices were motivated by the fact that l-C$_3$H/C$_4$H and C$^{34}$S/CCS belong to the same families of unsaturated carbon chains and should exhibit similar chemistries. The case of HC$_5$N is special as we have only one temperature determination. Since this temperature belongs to the warmest source in the best constrained carbon chain molecule, C$_4$H, it does not seem appropriate to simply adopt it in the sample. Nor does it seem appropriate to adopt the HC$_3$N or C$_4$H median temperatures however, since in SVS 4-5 the HC$_5$N temperature is 3 and 4 times higher, respectively. We instead opt to scale the SVS 4-5 HC$_5$N temperature with the ratio of the SVS 4-5 / sample median temperature for C$_4$H. The resulting temperature of 28~K is used to calculate HC$_5$N column densities and upper limits for all sources in the sample. The resulting estimates should be considered order of magnitude constraints. If instead, we adopt a HC$_5$N rotational temperature of 9~K, the lowest among the molecules, typical HC$_5$N column densities increase by two orders of magnitude, while the distribution of column densities is relatively unchanged. The choice of excitation temperature thus affects any conclusions on the relative abundances of HC$_5$N with respect to other molecules, but not conclusions based on correlations.

Since several protostars such as IRAS 03254$+$3050 and B1-c displayed moderate self-absorption in CS lines, we used the isotopologue ratios (CS/C$^{34}$S and C$^{34}$S/C$^{33}$S) to check the optical depth for each protostar. The integrated intensity ratios of CS/C$^{34}$S are between 5.0 and 20.5 and are all lower than the expected isotopic ratio of ${\sim}22$ \citep{Vocke09}, albeit marginally so for sources IRAS 03271 and L1448 IRS1. Therefore, we conclude that CS is optically thick in almost all sources. Next, we compared C$^{34}$S to C$^{33}$S and found ratios ranging from 4.9 to 9.3 in 15/16 sources, which, within uncertainties, are all consistent with the expected isotopic ratio of ${\sim}5.5$ \citep{Vocke09} and indicate that both lines are optically thin. The one exception is L1448 IRS1, where the ratio is 1.5. We used the C$^{34}$S line to calculate column densities for all sources with the caveat that we are certainly underestimating the column density for L1448 IRS1 (however, this source is not used for further analysis). Using Equation 3 from \citet{Takano98}, we find the highest value of optical depth for the C$^{34}$S line among our samples (excluding L1448 IRS1) was $\tau=0.072$ for source B1-a.

Across the sample, a few sources stand out for their especially rich or poor carbon chain inventories, or because they are enhanced in or lack specific molecules. The sources SVS 4-5 and B1-a have large carbon chain column densities compared to the other sources. All seven molecules are detected toward these sources and the derived column densities are often an order of magnitude in excess of the sample medians. In contrast, L1448 IRS1 and L1489 IRS have zero and two carbon chain detections, respectively. While HH 300 shows enhancement in sulfur-bearing molecules C$^{34}$S and CCS as well as carbon chain radical C$_4$H, no cyanpolyynes are observed toward this source. L1489 IRS also displays an enhancement in C$^{34}$S and CCS along with HC$_3$N, but lacks detections of the carbon chains l-C$_3$H and C$_4$H. 

Since only upper or lower limits could be constrained for all molecules in L1448 IRS1, we have excluded this source in all subsequent analysis. It is possible that its lack of carbon chains could be related to its high mass. Carbon chain deficiencies have been previously observed in massive YSOs \citep{Tercero10, Pagani17} and massive clumps in infrared dark clouds \citep{Sakai08}. In both cases, these deficiencies have been attributed to more advanced stages of chemical evolution compared to their lower-mass counterparts. It is unclear, however, whether this scenario can also explain the low abundances of unsaturated organics in L1448 IRS1 compared to the sample median.

\begin{figure*}[!htp]
\includegraphics[width=\linewidth]{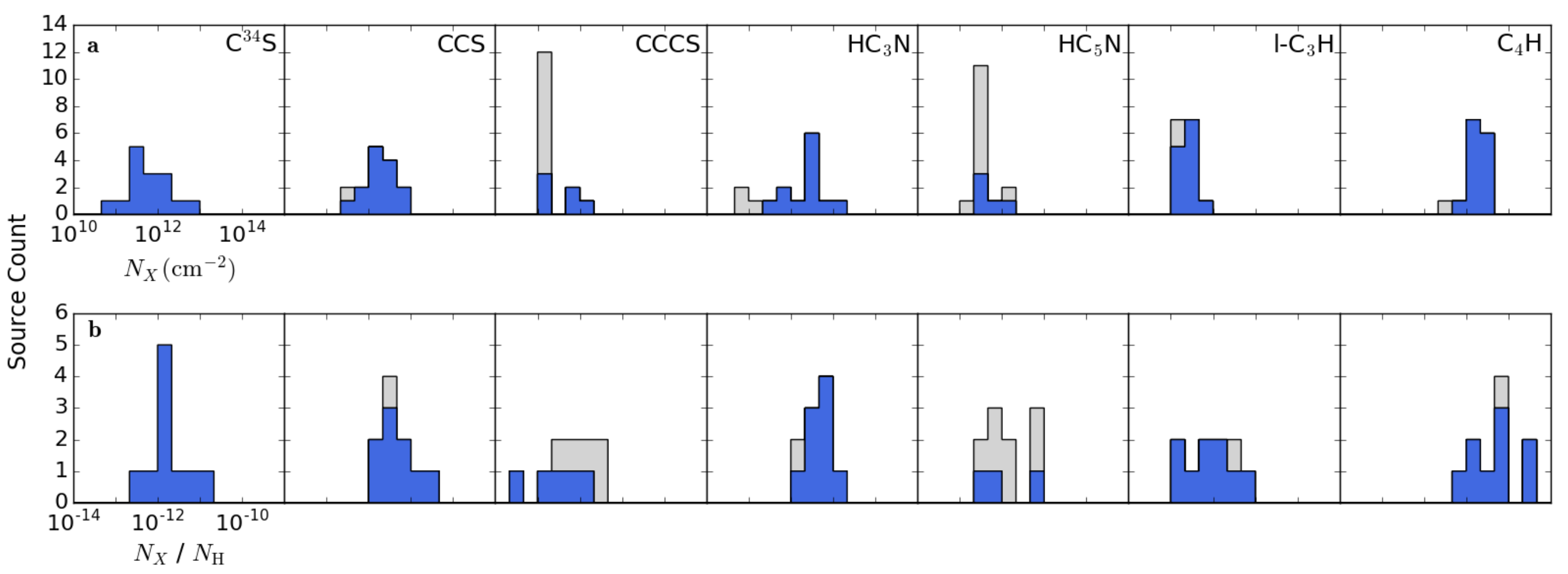}
\caption{a: Histograms of observed column densities for each molecule. b: Abundances with respect to hydrogen column density. In both panels, detections are shown in blue and upper limits in light gray. Tentative detections are shown in light gray, as are the upper limits.}
\label{fig:Molecules_Histograms}
\end{figure*}

The distributions of column densities and fractional abundances reveal variations in chemistry among sources and between different molecules. Figure \ref{fig:Molecules_Histograms}a shows a histogram of each molecule's column density across the sample. The column densities of C$^{34}$S, CCS, and HC$_3$N span two orders of magnitude or more. By comparison, the l-C$_3$H and C$_4$H distributions are narrow. Table \ref{tab:Median_Column_Densities} lists lower and upper quartiles for the column density of each molecule. For CCS and HC$_3$N, lower and upper quartile values differ by a factor of ${\sim}3$--$4$, while l-C$_3$H and C$_4$H quartiles differ by a factor of ${\sim}2$. Due to the large fraction of non-detections for CCCS and HC$_5$N, we are not able to reliably compare the distributions or quartile ranges of these species.

\begin{longrotatetable}
\begin{deluxetable*}{lccccccc}
\tablecaption{Carbon Chain Column Densities in cm$^{-2}$ for the 16 Sources\label{tab:column_density_large_table}}
\tablehead{[-.3cm]
Source & C$^{34}$S & CCS & CCCS & HC$_3$N & HC$_5$N & l-C$_3$H &  C$_4$H\tablenotemark{$a$} \\[-.55cm]}
\startdata
L1448 IRS1 & $>$ 8.7 $\thinmuskip=-3mu\times 10^{10,}$\tablenotemark{$b$} & $<$ 8.5$\thinmuskip=-3mu\times 10^{10}$ & $<$ 1.0$\thinmuskip=-3mu\times 10^{11}$ & $<$ 6.8$\thinmuskip=-3mu\times 10^{10}$ & $<$ 1.9$\thinmuskip=-3mu\times 10^{11}$ & $<$ 8.3$\thinmuskip=-3mu\times 10^{10}$ & $<$ 7.3$\thinmuskip=-3mu\times 10^{12}$ \\ 
IRAS 03235+3004 & 6.0 [2.9]$\thinmuskip=-3mu\times 10^{11}$& 2.2 [0.7]$\thinmuskip=-3mu\times 10^{12}$ & 2.0 [4.3]$\thinmuskip=-3mu\times 10^{11}$ & 3.7 [1.1]$\thinmuskip=-3mu\times 10^{12}$ & $<$ 2.7$\thinmuskip=-3mu\times 10^{11}$ & 5.5 [2.9]$\thinmuskip=-3mu\times 10^{11}$ & 3.8 [1.5]$\thinmuskip=-3mu\times 10^{13}$\\
IRAS 03245+3002 & 1.4 [0.7]$\thinmuskip=-3mu\times 10^{12}$ & 3.3 [1.0]$\thinmuskip=-3mu\times 10^{12}$ & 7.0 [2.5]$\thinmuskip=-3mu\times 10^{11}$ & 4.0 [2.1]$\thinmuskip=-3mu\times 10^{12}$ & 7.3 [10.8]$\thinmuskip=-3mu\times 10^{11}$ & 1.7 [0.9]$\thinmuskip=-3mu\times 10^{11}$ & 1.6 [0.7]$\thinmuskip=-3mu\times 10^{13}$\\
L1455 SMM1 & 1.4 [0.7]$\thinmuskip=-3mu\times 10^{12}$ & 4.3 [1.4]$\thinmuskip=-3mu\times 10^{12}$ & 1.8 [0.1]$\thinmuskip=-3mu\times 10^{11}$ & 2.7 [1.4]$\thinmuskip=-3mu\times 10^{12}$ & $<$ 2.5$\thinmuskip=-3mu\times 10^{11}$ & 3.5 [1.9]$\thinmuskip=-3mu\times 10^{11}$ & 1.8 [0.5]$\thinmuskip=-3mu\times 10^{13}$\\
L1455 IRS3 & 4.2 [2.0]$\thinmuskip=-3mu\times 10^{11}$ & 1.4 [0.2]$\thinmuskip=-3mu\times 10^{12}$ & $<$ 1.1$\thinmuskip=-3mu\times 10^{11}$ & 7.3 [3.8]$\thinmuskip=-3mu\times 10^{11}$ & $<$  2.6$\thinmuskip=-3mu\times 10^{11}$ & 2.5 [1.4]$\thinmuskip=-3mu\times 10^{11}$ & 1.1 [0.3]$\thinmuskip=-3mu\times 10^{13}$\\
IRAS 03254+3050 & 7.6 [4.0]$\thinmuskip=-3mu\times 10^{10}$ & $\mathit{2.3\,[1.8]\thinmuskip=-3mu\times 10^{11}}$ & $<$ 1.5$\thinmuskip=-3mu\times 10^{11}$ & $<$ 9.3$\thinmuskip=-3mu\times 10^{10}$ & 3.9 [4.1]$\thinmuskip=-3mu\times 10^{11}$ & 1.9 [1.1]$\thinmuskip=-3mu\times 10^{11}$ &  1.4 [0.9]$\thinmuskip=-3mu\times 10^{13}$ \\
IRAS 03271+3013 & 1.1 [0.6]$\thinmuskip=-3mu\times 10^{11}$& 3.0 [2.4]$\thinmuskip=-3mu\times 10^{11}$ & $<$ 1.1$\thinmuskip=-3mu\times 10^{11}$ & 1.9 [1.0]$\thinmuskip=-3mu\times 10^{12}$ & $<$ 2.8$\thinmuskip=-3mu\times 10^{11}$ & 4.3 [2.3]$\thinmuskip=-3mu\times 10^{11}$ & 1.4 [0.5]$\thinmuskip=-3mu\times 10^{13}$ \\
B1-a & 3.6 [1.7]$\thinmuskip=-3mu\times 10^{12}$ & 5.8 [1.7]$\thinmuskip=-3mu\times 10^{12}$ & 6.1 [4.9]$\thinmuskip=-3mu\times 10^{11}$ & 4.1 [1.2]$\thinmuskip=-3mu\times 10^{12}$ & $\mathit{2.5\,[2.6]\thinmuskip=-3mu\times10^{11}}$ & 3.9 [2.1]$\thinmuskip=-3mu\times 10^{11}$ & 2.5 [0.6]$\thinmuskip=-3mu\times 10^{13}$ \\
B1-c & 1.3 [0.6]$\thinmuskip=-3mu\times 10^{12}$ & 6.2 [2.9]$\thinmuskip=-3mu\times 10^{12}$ & 1.3 [4.9]$\thinmuskip=-3mu\times 10^{11}$ & 4.7 [2.4]$\thinmuskip=-3mu\times 10^{12}$ & 1.2 [1.6]$\thinmuskip=-3mu\times 10^{12}$ & 4.3 [2.3]$\thinmuskip=-3mu\times 10^{11}$ & 2.7 [0.2]$\thinmuskip=-3mu\times 10^{13}$\\
B5 IRS 1 & 8.8 [4.3]$\thinmuskip=-3mu\times 10^{11}$ & 1.7 [0.5]$\thinmuskip=-3mu\times 10^{12}$  & $<$ 1.8$\thinmuskip=-3mu\times 10^{11}$ & 3.2 [1.0]$\thinmuskip=-3mu\times 10^{12}$ & $\mathit{1.0\,[1.3]\thinmuskip=-3mu\times 10^{12}}$ & 4.0 [2.3]$\thinmuskip=-3mu\times 10^{11}$ & 3.6 [0.5]$\thinmuskip=-3mu\times 10^{13}$\\
L1489 IRS & 2.5 [1.3]$\thinmuskip=-3mu\times 10^{11}$ & 5.4 [5.2]$\thinmuskip=-3mu\times 10^{11}$ & $<$ 1.1$\thinmuskip=-3mu\times 10^{11}$ & 3.8 [3.9]$\thinmuskip=-3mu\times 10^{11}$  & $<$  4.3$\thinmuskip=-3mu\times 10^{11}$ & $<$ 1.3$\thinmuskip=-3mu\times 10^{11}$ & $<$ 3.2$\thinmuskip=-3mu\times 10^{12}$\\
IRAS 04108+2803 & 4.1 [2.0]$\thinmuskip=-3mu\times 10^{11}$ & 6.0 [4.7]$\thinmuskip=-3mu\times 10^{11}$ & $<$ 1.5$\thinmuskip=-3mu\times 10^{11}$ & $<$ 6.0$\thinmuskip=-3mu\times 10^{10}$ & $<$ 2.4$\thinmuskip=-3mu\times 10^{11}$ & 1.9 [1.1]$\thinmuskip=-3mu\times 10^{11}$ & 6.4 [2.4]$\thinmuskip=-3mu\times 10^{12}$\\
HH 300 & 4.4 [2.2]$\thinmuskip=-3mu\times 10^{11}$ & 1.2 [0.8]$\thinmuskip=-3mu\times 10^{12}$ & $<$ 1.7$\thinmuskip=-3mu\times 10^{11}$ & $<$ 1.4$\thinmuskip=-3mu\times 10^{11}$ & $<$ 3.4$\thinmuskip=-3mu\times 10^{11}$ & $<$ 1.4$\thinmuskip=-3mu\times 10^{11}$ & 1.0 [3.2]$\thinmuskip=-3mu\times 10^{13}$ \\
SVS 4-5 & 4.9 [2.4]$\thinmuskip=-3mu\times 10^{12}$& 3.5 [1.3]$\thinmuskip=-3mu\times 10^{12}$ & 1.4 [1.7]$\thinmuskip=-3mu\times 10^{12}$ & 1.0 [0.3]$\thinmuskip=-3mu\times 10^{13}$ & 2.6 [1.2]$\thinmuskip=-3mu\times 10^{11}$ & 4.1 [2.2]$\thinmuskip=-3mu\times 10^{11}$ & 2.2 [0.4]$\thinmuskip=-3mu\times 10^{13}$\\
L1014 IRS & 2.9 [1.4]$\thinmuskip=-3mu\times 10^{11}$ & 2.1 [0.8]$\thinmuskip=-3mu\times 10^{12}$ & $<$ 1.1$\thinmuskip=-3mu\times 10^{11}$ & 4.9 [2.5]$\thinmuskip=-3mu\times 10^{11}$ & $<$ 1.7$\thinmuskip=-3mu\times 10^{11}$ & 2.0 [1.1]$\thinmuskip=-3mu\times 10^{11}$ & 3.2 [0.8]$\thinmuskip=-3mu\times 10^{13}$\\
IRAS 23238+7401 & 7.5 [3.7]$\thinmuskip=-3mu\times 10^{11}$ & 1.7 [0.2]$\thinmuskip=-3mu\times 10^{12}$ & $\mathit{1.5\,[3.3]\thinmuskip=-3mu\times 10^{11}}$ & 3.6 [1.8]$\thinmuskip=-3mu\times 10^{12}$ & 3.7 [3.7]$\thinmuskip=-3mu\times 10^{11}$ & 2.1 [1.2]$\thinmuskip=-3mu\times 10^{11}$ & 1.4 [2.2]$\thinmuskip=-3mu\times 10^{13}$\\ \hline
Species Median & 6.0$^{13.5}_{3.5}\thinmuskip=-3mu\times 10^{11}$ & 1.7$^{3.4}_{0.90}\thinmuskip=-3mu\times 10^{12}$ & 2.0$_{1.7}^{6.6}\thinmuskip=-3mu\times 10^{11}$ & 3.4$^{4.0}_{1.6}\thinmuskip=-3mu\times 10^{12}$ & 3.9$^{8.7}_{3.2}\thinmuskip=-3mu\times 10^{11}$ & 3.5$_{2.0}^{4.1}\thinmuskip=-3mu\times 10^{11}$ & 1.7$^{2.7}_{1.4}\thinmuskip=-3mu\times 10^{13}$ \\ 
\enddata
\tablenotetext{a}{C$_4$H data are taken from \citet{Graninger16} with new detection from source IRAS 03254$+$3050.\\[-.4cm]}
\tablenotetext{b}{Due to significant self-absorption in this source, we are only able to constrain a lower limit.\\[-.4cm]}
\tablecomments{Uncertainties at the $1\sigma$ level are reported in brackets and tentative detections are shown in italics. Species median column densities with quartiles are reported using only calculated column densities (i.e.,~no upper or lower limits). While we list the column density data for L1448 IRS1, we exclude this source in all subsequent analysis.}
\end{deluxetable*}
\end{longrotatetable}

\subsection{Fractional Abundance}

To compare our observational results with theoretical predictions for low-mass YSO chemical abundances, we need to calculate fractional abundances with respect to atomic hydrogen. Thus, we first need to determine the hydrogen column density that is contained within the beam. To obtain an order-of-magnitude estimate (more precise estimates require detailed models of each individual source, which is beyond the scope of this study), we use existing constraints on envelope masses toward the majority of our sources and assume a spherically symmetric physical model with power-law density profile:

\begin{equation}
n_{\rm{H}} (r) = n_{\rm{M_{env}}} \left( \frac{r}{1000\,\rm{AU}} \right)^{-\alpha}
\end{equation}

\noindent where $n_{\rm{M_{env}}}$ is a normalizing constant based on the protostellar envelope mass, which we have estimates of for 11 sources (see Table \ref{tab:source_info}). We assume that all of the envelope material is molecular hydrogen. From the median values derived from radiative transfer modeling of low-mass protostars by \citet{Jorgensen02}, $\alpha=1.5$, which is also the expected value of a free-falling core. To determine the total number of H nuclei within the beam for a typical protostar with radial density profile $n_{\rm{H}} (r)$ and emitting area $A(r)$, we integrate radius from minimum ($r_{\rm{min}}$) to maximum ($r_{\rm{max}}$) :

\begin{equation}
\eta_{\rm{H}} = \displaystyle \int_{r_{\rm{min}}}^{r_{\rm{max}}} A(r) n_{\rm{H}}(r) dr.
\end{equation}

\noindent For our sources, we assume $r_{\rm{min}}=1\,\rm{AU}$ and a maximum extent of $r_{\rm{max}} = 20000\,\rm{AU}$, which is consistent with the values adopted in \citet{Jorgensen02}. We can express $A(r)$, in terms of the cylindrical line of sight, as:

\begin{eqnarray}
A(r)=\left\{ \begin{array}{ll}
\renewcommand{\arraystretch}{3}
4 \pi r^2, & \,\,\,\, r\leq r_b \\
2\pi \left[r_b^2 + \left(r - \sqrt{r^2 - r_b^2} \right)^2 \right], & \,\,\,\, r>r_b. \\
\end{array}
\right.
\end{eqnarray}

Equivalently, the emitting surface, when the radius is less than the beam radius, is that of the surface area of a sphere at that particular radius. For radii larger than the beam radius, the emitting surface corresponds to that of spherical caps in front of and behind a sphere with radius $r_b$. Each cap has a fixed base radius equal to the beam radius $r_b$ and a height that is allowed to vary based on radius. This methodology is adapted from \citet{Bergner17}.

Using this method, the beam is determined to contain $\approx\,$50\% and $\approx\,$46\% of the protostellar envelope mass for Perseus and Taurus sources, respectively. To test the effect of selecting different $r_{\rm{max}}$ values, beam fractions were calculated for $r_{\rm{max}}=10000\,\rm{AU}$. For Perseus sources, there was a difference of about 8\% in beam fraction and about 5\% for Taurus sources.

Then, the beam-averaged hydrogen column density is given by:

\begin{equation}
N_{\rm{H}} = \frac{\eta_{\rm{H}}}{\pi r_b^2}
\end{equation}

\noindent where $r_b$ is the beam radius. We adopt a $12^{\prime \prime}$ beam radius based on the median observing beam size (see Section \ref{sec:observations}), and a distance of 250~pc to Perseus \citep{Schlafly14} and 140~pc to Taurus \citep{Torres07}. 

Given the column density $N_X$ of a species, the species specific fractional abundance is simply given by the ratio:

\begin{equation}
f_X = \frac{N_X}{N_{\rm{H}}}.
\end{equation}

Fractional abundance distributions are shown in Figure \ref{fig:Molecules_Histograms}b. Since envelope masses are not available for SVS~4-5, IRAS~23238, L1014~IRS, HH~300, and IRAS~04108, these sources are omitted from the comparison. The spread in carbon chain abundances is considerable, demonstrating that the column density spread (as seen in Figure \ref{fig:Molecules_Histograms}a) is not simply an effect caused by different source sizes. Relative to the column densities, the abundance distributions broaden for l-C$_3$H and C$_4$H, tighten for HC$_3$N, and remain roughly the same for C$^{34}$S and CCS. This tightening of the HC$_3$N distribution is particularly prominent, which in terms of column density, had previously spanned two orders of magnitude but in fractional abundance only spans a little over one order of magnitude.

The calculated fractional abundances are naturally quite uncertain, since in reality, envelope sizes and radial profiles differ substantially between sources \citep[e.g.,][]{Jorgensen02}. We note that all sources but one for which this analysis is possible (i.e., envelopes masses are known) are located in Perseus, which implies that we do not suffer from errors related to different relative distances within the sample. Still, better constraints on both envelope and carbon chain radial profiles are required for meaningful data-model comparisons beyond order-of-magnitude checks.

\subsection{Median Column Densities and Abundances}

We calculate median column densities and abundances using both detections only and detections with upper limits. If median abundances are determined by using only detections, we are omitting information supplied by the upper limits and risk over-estimating median occurrences. Hence, we used survival analysis in the form of the Kaplan-Meier (KM) estimator with left censorship \citep{Feigelson85}. All detections and non-detections were ordered and then the values of positive detections were used to define intervals for the total range of values. Whenever an upper limit was contained in an interval, it was treated as being the lower limiting value of that interval. Equivalently, every positive detection was weighted by the number of upper limits between it and the next largest positive detection.

\begin{figure*}[!htp]
\includegraphics[width=\linewidth]{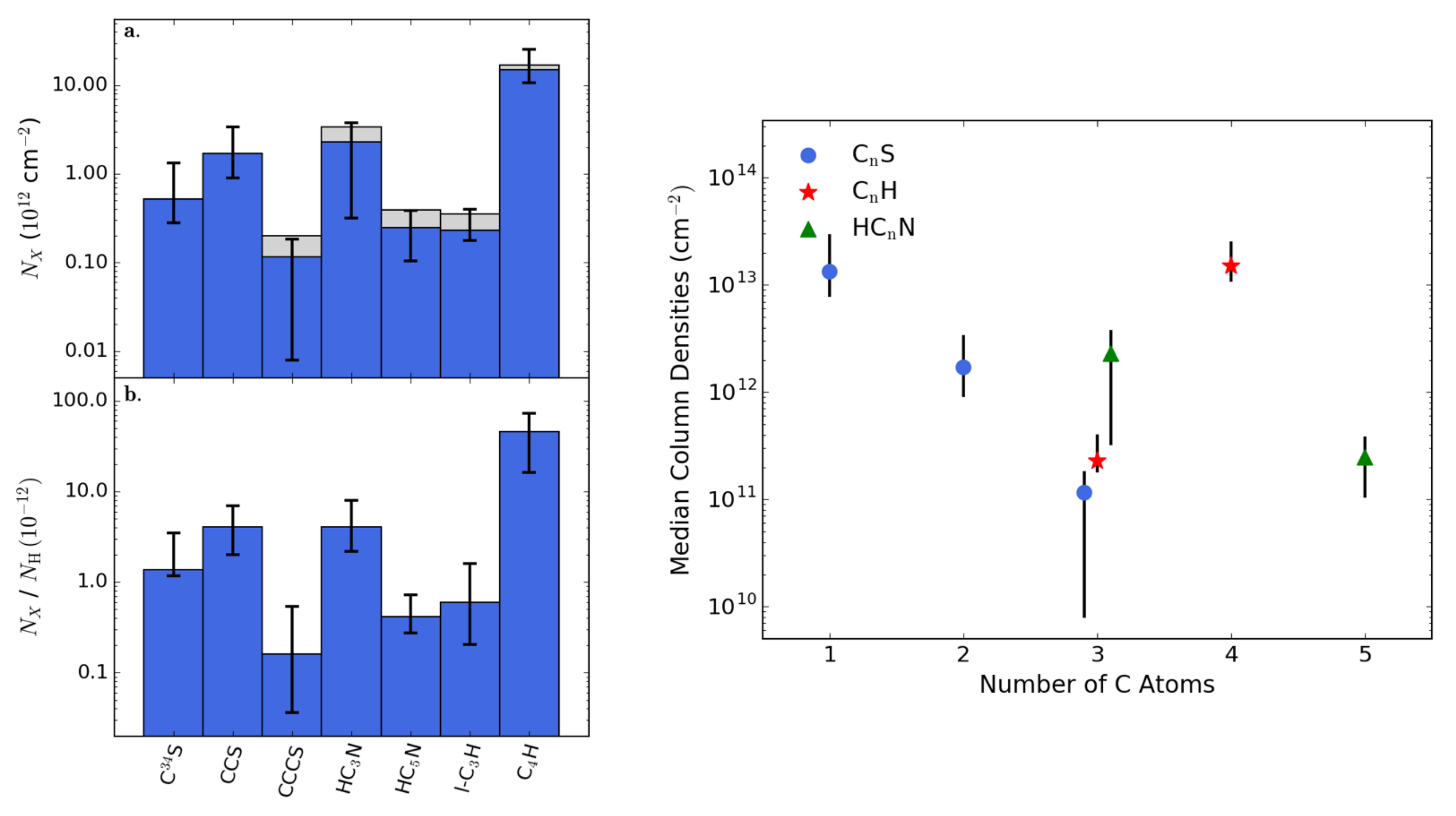}
\caption{\textit{Left:} a: Median column densities, calculated with detections only (light gray) and with survival analysis (dark blue). b: Median fractional abundances and error bars spanning the first and third quartiles calculated with survival analysis. \textit{Right:} Number of carbon atoms versus median column densities. Error bars span the first and third quartile derived by survival analysis. As they each contain three carbon atoms, CCCS, l-C$_3$H, and HC$_3$N have small horizontal offsets for the sake of visual clarity. The median column density for C$^{34}$S was multiplied by 22 to get the CS median column.}
\label{fig:Median_ColumnDensities}
\end{figure*}

To account for the discrete nature of the survival function, we calculated median abundances via linearly interpolating between the values above and below where the cumulative density function (CDF) equaled 0.5. But, if the lowest 50\% of values consist of solely upper limits, the first positive detection will happen after the CDF has already exceeded 0.5 and it will be impossible to compute medians. In our sample, this applies to CCCS and HC$_5$N. To circumvent this limitation, we assigned the lowest value in our sample a ``detection" status, irrespective of its true nature as a detection or a non-detection and then calculated medians via the KM estimator \citep[e.g.,][]{Feigelson85, Bergner17}. While this method may artificially inflate estimates for median values, we believe this procedure still provides a more realistic estimate of the median than simply using detections alone. As can be seen in Figure \ref{fig:Median_ColumnDensities} and Table \ref{tab:Median_Column_Densities}, there is a significant difference between medians determined using only detections and those calculated with survival analysis, which reveals the importance of incorporating the constraints provided by non-detections. Note that we include tentative detections for CCS, CCCS, and HC$_5$N to estimate their median column densities. The inclusion of tentative detections does not have a large effect on the median column densities (${\sim}10$\%) but increases the HC$_5$N median fractional abundance by ${\sim}50$\%.

\begin{deluxetable*}{lccc}[!htp]
\tablecaption{Median Column Densities and Fractional Abundances\label{tab:Median_Column_Densities}}
\tablehead{[-.3cm]
& \multicolumn{2}{c}{Median Column Densities} & Median Abundances \\
& \multicolumn{2}{c}{($10^{12}\,\rm{cm}^{-2}$)} & $\left(10^{-12}\right)$\\
& Detections Only & Survival Analysis  & Survival Analysis} 
\startdata
C$^{34}$S & $0.60_{0.35}^{1.35}$  & $0.60_{0.35}^{1.35}$  & $1.38_{1.17}^{3.48}$ \\ 
CCS & $1.7^{3.4}_{0.90}$  & $1.7^{3.4}_{0.90}$ & $4.08_{2.01}^{6.95}$ \\
CCCS & $0.20_{0.17}^{0.66}$ & $0.12_{0.0079}^{0.19}$ & $0.16_{0.037}^{0.54}$ \\
HC$_3$N & $3.4_{1.6}^{4.0}$ & $2.3_{0.32}^{3.8}$ & $4.06_{2.18}^{8.00}$ \\
HC$_5$N & $0.39_{0.32}^{0.87}$ & $0.25_{0.10}^{0.39}$ & $0.41_{0.28}^{0.72}$ \\ 
l-C$_3$H & $0.35_{0.20}^{0.41}$ & $0.23_{0.18}^{0.40}$ & $ 0.60_{0.20}^{1.61}$  \\
C$_4$H & $17.0_{14.0}^{26.5}$  & $15.0_{10.7}^{25.5}$ &  $46.0_{16.2}^{73.2}$ \\
\enddata
\tablecomments{Lower and upper quartiles are shown to the right of the medians.}
\end{deluxetable*}

Median column densities span three orders of magnitude. For the sulfur-bearing chains, each additional carbon atom leads to approximately an order of magnitude reduction in the median column density. Similarly, the addition of two C atoms from HC$_3$N to HC$_5$N leads to a reduction of one order of magnitude. For the carbon chain radicals, however, the addition of one C atom from l-C$_3$H to C$_4$H leads to a two order of magnitude increase in column density.

Median fractional abundances calculated using survival analysis are shown in Figure \ref{fig:Median_ColumnDensities}b. The median fractional abundances also span three orders of magnitude and exhibit the same trends as found for the median column densities. The distribution of abundances, reflected in the error bars spanning the lower and upper quartiles, is tighter for the small sulfur-bearing molecules C$^{34}$S and CCS than for both the nitrogen-bearing chains and the carbon chain radicals.

\subsection{Correlation Studies}

To assess chemical relationships between carbon chains, we searched for correlations between different species. Both chemically-related molecules (e.g,.~if one forms from another) or those that depend similarly on an underlying physical parameter (e.g.,~envelope mass, temperature) are expected to exhibit strong positive correlations. When comparing column densities, some correlation is expected regardless of chemical relationships, as most molecules increase with an increasing total column density along a sight line. In fact, we see a positive correlation between column densities and envelope masses for all molecules, except CCCS (not shown). We therefore focus our correlation study on the derived abundances.

\begin{figure*}[!htp]
\centering
\includegraphics[width=\linewidth]{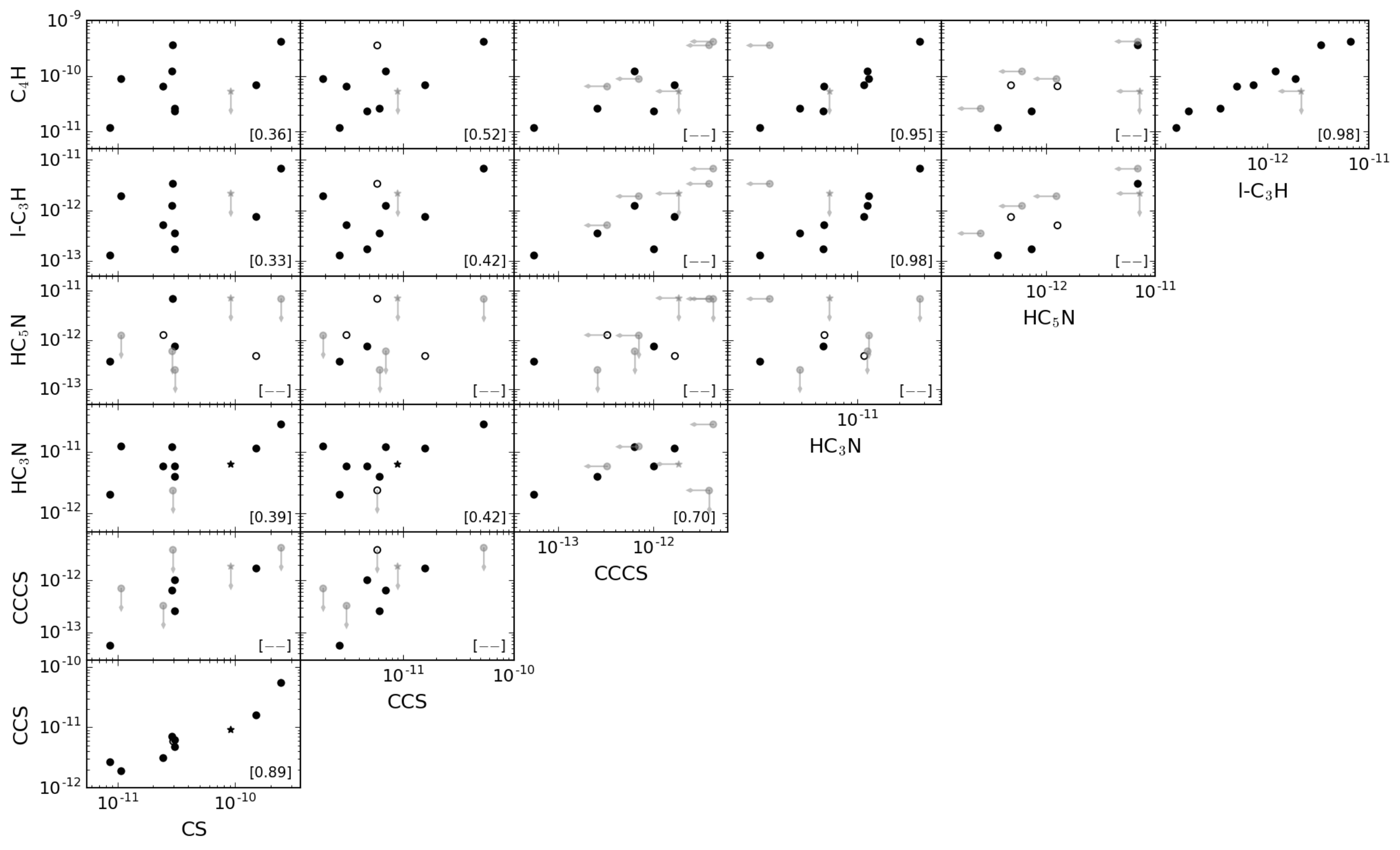}
\caption{Fractional abundances of each molecule plotted against one another. Detections are shown in black and upper limits as light gray arrows, while tentative detections are shown as open circles. Sources which are members of Perseus are shown as circles and all other sources are represented by stars. Spearman correlation coefficients using all detected and tentatively-detected sources are displayed in the bottom right corners of each scatter plot. As correlation coefficients with $\leq$ 5 mutual detections are not robust, these correlations, namely those involving the species CCCS and HC$_5$N, are denoted by dashes (`$--$').}
\label{fig:Correlation_Matrix_Column_Densities_Plot}
\end{figure*}

We calculate the Spearman's rank correlation coefficient for each carbon chain with one another, as shown in Figure \ref{fig:Correlation_Matrix_Column_Densities_Plot} and list the results in Table \ref{tab:spear_table_results}. Unlike Pearson's correlation, which characterizes linear relationships, Spearman's correlation assesses monotonic relationships or alternatively, the rank order between two variables. Spearman's correlation was chosen since we do not have any prior expectation of linear correlation and because Spearman's correlation is less sensitive to outlier values than Pearson's. Only detections were used for calculating the correlations with the implication that for some species, the relatively low number of mutual detections implies that our determined correlation coefficients are not robust to the inclusion/exclusion of a few low- or high-valued data points. We also note that, if we instead consider Pearson's correlations, all identified correlations are still found to be significant.

Correlations are found at the 90\% confidence level between similar families of carbon chains (e.g., CS/CCS, l-C$_3$H/C$_4$H) and between HC$_3$N and the carbon chains l-C$_3$H and C$_4$H. This latter correlation could be explained by a closely-related production chemistry of l-C$_3$H and C$_4$H and that of HC$_3$N in low-mass protostars, as discussed in Section \ref{sec:CC_chemistry}. This result is independent of whether we only consider Perseus sources or also include the one Taurus source for which we could calculate abundances. However, the correlation could also be due to a mutual dependence on an unobserved underlying variable. Without a clear mechanism or a multivariable correlation study, these two explanations cannot be distinguished.

\begin{deluxetable*}{lcccccc}[!htp]
\tablecaption{Correlation Coefficients for Molecular Abundance Correlations\label{tab:spear_table_results}}
\tablehead{[-.3cm]
& CS & CCS  & HC$_3$N &  l-C$_3$H  \\[-.55cm] }
\startdata
C$_4$H   & 0.36 [9] & 0.52 [9] &  \textbf{0.95 [8]} & \textbf{0.98 [9]} \\
l-C$_3$H   & 0.33 [9] & 0.42 [9] &  \textbf{0.98 [8]} \\
HC$_3$N & 0.39 [9] & 0.42 [9] &  \\
CCS         & \textbf{0.89 [10]} & & \\ 
\enddata
\tablecomments{Brackets represent the number of sources with detections, including those that are tentative, for both molecules. Correlations that exceed the 90\% confidence level are shown in bold. Correlations coefficients for CCCS and HC$_5$N are omitted due to the small number ($\leq$ 5) of mutual detections.}
\end{deluxetable*}

\begin{figure*}[!htp]
\centering
\includegraphics[width=\linewidth]{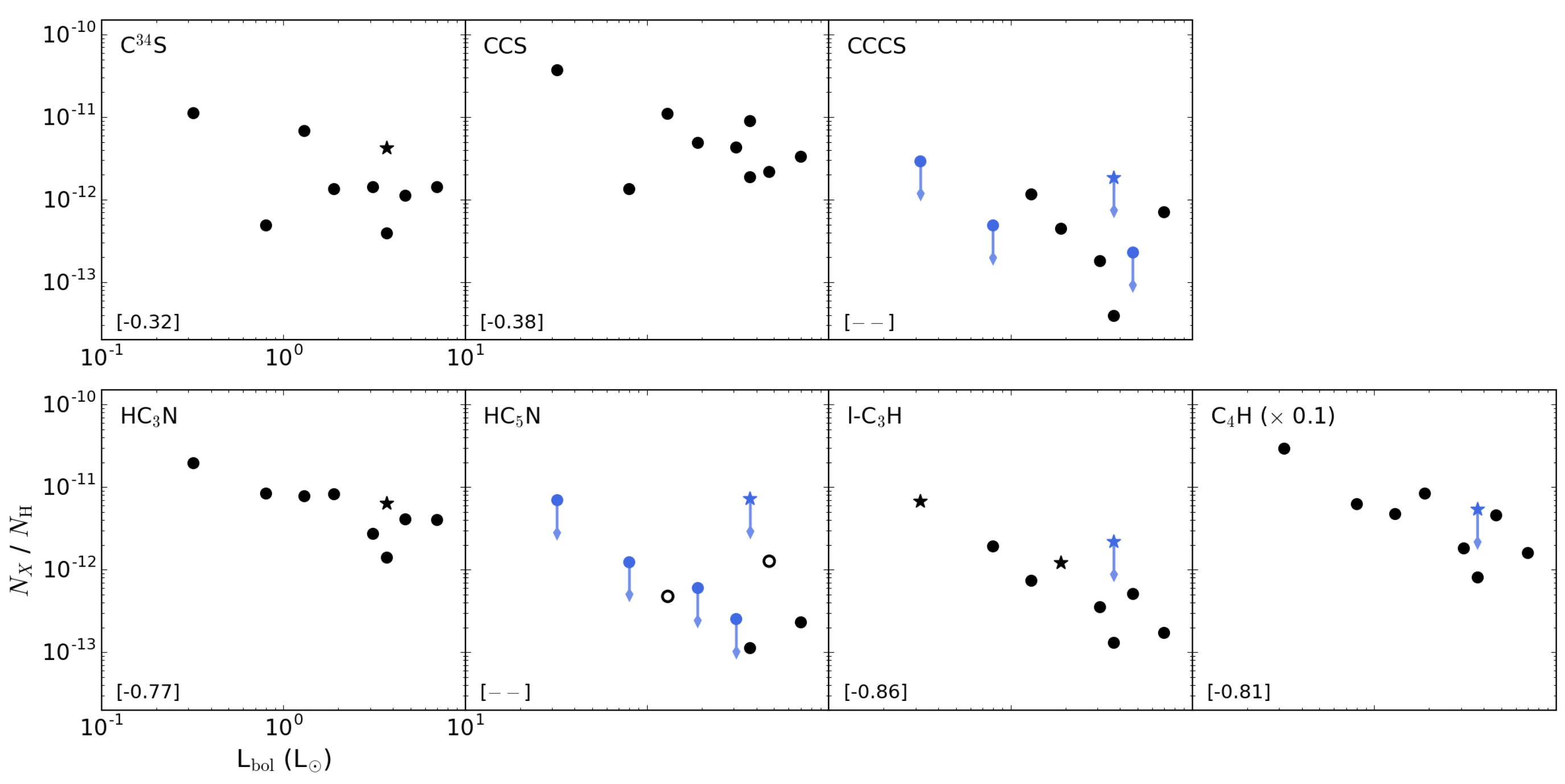}
\caption{Fractional abundances plotted against bolometric luminosity. Upper limits due to non-detections are indicated by blue arrows and tentative detections are shown as open circles. Sources which are members of Perseus are shown as circles and all other sources are represented by a star. Spearman correlation coefficients using all detected and tentatively-detected sources are shown in brackets in the lower right corner. For the sake of visual clarity, the column densities for C$_4$H are scaled by 0.1.}
\label{fig:Lum_Cor_Plot}
\end{figure*}

Correlations of abundances with physical properties may also yield insight into the underlying chemistry, e.g. if the carbon chemistry depends on protostellar heating, a positive correlation with protostellar luminosity. We checked for correlations between bolometric luminosity and carbon chain abundance and found moderate to strong negative correlations for all species. The luminosity correlations are presented in Figure \ref{fig:Lum_Cor_Plot}. Both l-C$_3$H and C$_4$H as well as HC$_3$N stood out as particularly well anti-correlated with bolometric luminosity. While this finding suggests that more luminous and evolved protostars have envelopes exhibiting lower carbon chain abundances, we find that most of this anti-correlation might be explained by a positive correlation ($r=0.69$) between hydrogen column density and bolometric luminosity.

\section{Discussion}
\label{sec:discussion}

\subsection{Carbon Chain Chemistry}
\label{sec:CC_chemistry}

The observed carbon chain abundances provide constraints on the chemistry that could produce them. We note that such a chemistry must be general enough to always produce some carbon chains, as we detect them in almost all sources. The formation and/or destruction chemistry must also be sensitive to some aspects of the protostellar environment and/or evolutionary stage since we see order-of-magnitude variations across the sample. As shown in the right panel of Figure \ref{fig:Median_ColumnDensities}, the decreasing column density with size in the cyanpolyynes and sulfur-bearing chains is consistent with bottom-up chemistry, as is the increasing column density from l-C$_3$H to C$_4$H, since these molecules form via different pathways (see subsequent discussion). Finally, the lack of correlations between the pure hydrocarbon chains and the sulfur-bearing molecules suggest that there are multiple, fairly independent carbon chain chemistries that affect the final carbon chain composition in a protostellar envelope. 


Within the pure hydrocarbon chain molecular family, we find that the relative abundances of l-C$_3$H and C$_4$H are consistent with model expectations \citep{Hassel11} and previous observations \citep{Gratier16}. The fact that C$_{\rm{n}}$H with even-numbered n tends to be much more abundant than species with odd n is well-known and is related to the molecular structure and reactivity of these species \citep[e.g.,][]{Cernicharo87, Kawaguchi91, Guelin97, Sakai10}. Based on our observations, we find that this trend is also present in the relatively warm and dense regions of gaseous envelopes surrounding protostars. C$_4$H is expected to form efficiently at low temperatures \citep{Chastaing98} through two different pathways involving $\rm{C}_{2n}\rm{H}_2$ with C$_2$ \citep{Canosa07} and C$_2$H \citep{Agundez17}, respectively. The production of unsaturated carbon chains with an odd number of carbon atoms also occurs efficiently at low temperatures and is due to reactions of neutral carbon atoms and CH radicals with C$_2$H$_2$ \citep{Chastaing01, Loison09}. In particular, l-C$_3$H can form by the neutral-neutral reaction $\rm{C} + \rm{C}_2\rm{H}_2 \rightarrow \rm{C}_3\rm{H} + \rm{H}$ \citep{Kaiser97, Kaiser99}. \citet{Sakai08} also suggested a lukewarm pathway involving methane and C$^+$. The strong correlation between l-C$_3$H and C$_4$H is consistent with their shared origin in C$_2$H$_2$. It would be interesting to obtain more sensitive data and explore whether this correlation continues to be as strong when considering larger hydrocarbon chains. 

For the cyanopolyynes, we also find that the relative abundances of HC$_3$N and HC$_5$N are consistent with model predictions and previous observations \citep{Hassel11, Gratier16}. Cyanopolyynes can form efficiently at low temperatures \citep{Seki96} and without activation barriers \citep{Fukuzawa97, Choi04} via a neutral-neutral reaction of CN with $\rm{C}_{2\rm{n}}\rm{H}_2$. Additional formation pathways involve the growth of cyanopolyyne carbon chains by C$_2$H$_2^+ +$ HC$_{2\rm{n}+1}$N and reactions between hydrocarbon ions and nitrogen atoms followed by electron recombination reactions \citep[see Figure 2 in ][]{Taniguchi16}. Based on observational results, \citet{Taniguchi16} suggest that the hydrocarbon ion and N atoms reactions are the dominant pathway for forming HC$_5$N, while the neutral-neutral reaction is likely the most important for HC$_3$N \citep{Takano98}. The strong correlation we observe between HC$_3$N and the hydrocarbons is consistent with their shared origin in C$_2$H$_2$ and a neutral-neutral production pathway and is compatible with the above model. 

The decreasing trend in median column density from $n=1$--$3$ seen in the C$_{\rm{n}}$S molecules is consistent with previous observations and model expectations \citep{Hassel11, Gratier16}. Since the ionization potential of the sulfur atom is lower than that of the carbon atom, most of the sulfur atoms are ionized by the interstellar UV radiation in regions where there is only partial ionization of carbon atoms \citep{Suzuki92}. Since S$^+$ does not interact with lower vibrational states ($v \le 1$) of H$_2$ in cold clouds \citep{Prasad82, Zanchet13}, the S$^+$ ion is expected to be abundant in the gas phase in regions that produce carbon chains \citep{Suzuki92}. Reactions of hydrocarbons such as CH, C$_2$H, and l-C$_3$H with S$^+$ are major routes to produce CS, CCS, and CCCS, respectively \citep[see Figure 11 in ][]{Suzuki92}. Numerous neutral-neutral reactions have also been identified \citep[e.g.,][]{Yamada02, Sakai07, Agundez13} and may significantly contribute to the production of CCS and CCCS. In particular, \citet{Sakai07} found that the reaction between CH and CS is the most probable neutral-neutral pathway to produce CCS, while the formation of CCCS via the neutral-neutral reaction C$_2 +$ H$_2$CS $\to$ CCCS $+$ H$_2$ has been suggested by \citet{Yamada02}.

The strong mutual correlations between CS and CCS are consistent with both the proposed ion-molecule and neutral-neutral production routes. However, as our survey did not detect CCCS in the majority of sources, higher resolution data of sulfur-bearing chains and the pure hydrocarbons could help clarify the relative contributions of the proposed formation pathways of sulfur containing carbon chains, especially larger ($n\ge3$) chains, and to determine if neutral-neutral or ion-molecule reactions are more dominant in protostellar envelopes. Specifically, an observed strong correlation between CCCS and C$_3$H would support formation via ion-molecule reactions, while the absence of such a correlation would imply that neutral-neutral reactions are more important for CCCS production.

\subsection{Comparison with Cold Clouds}

\begin{figure}[!htp]
\includegraphics[width=\linewidth]{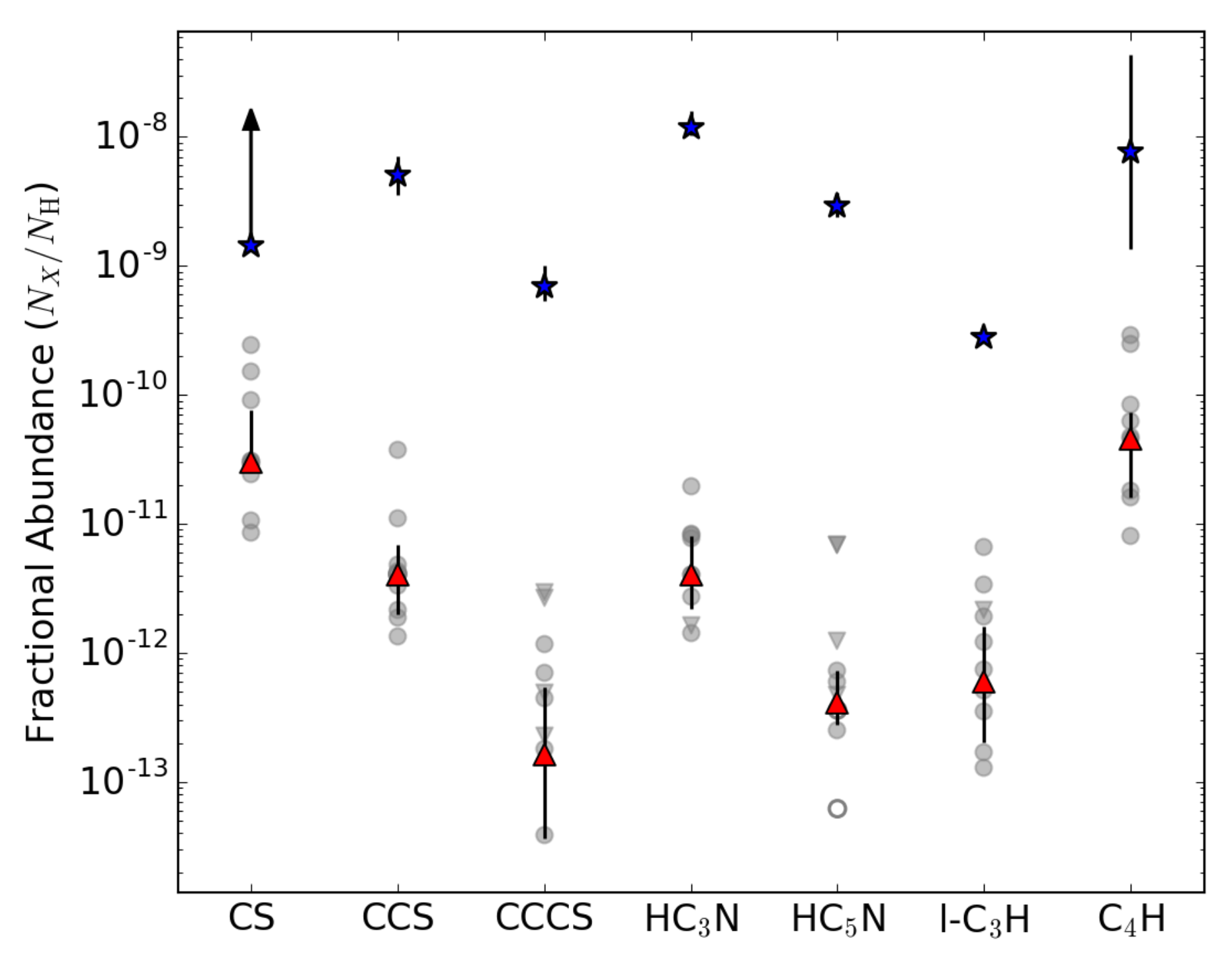}
\caption{Observed fractional abundances compared with cold cloud TMC-1. Median fractional abundances are represented by red upward triangles and error bars span the first and third quartiles derived by survival analysis. Individual sources are represented by gray circles for detections and downward triangles for upper limits. Blue stars indicate TMC-1 observed fractional abundances from \citet{Gratier16}.}
\label{fig:ColdCloud_Corino_Comparison}
\end{figure}

We compare our carbon chain abundances against those found in dense cold clouds to address questions of chemical inheritance. Due to its relative closeness (${\sim}140\,\rm{pc}$) and rich molecular complexity, TMC-1 has been extensively observed to study the chemical processes taking place in dark clouds \citep[e.g.,][]{Ohishi98, Kaifu04, Smith04, McCarthy06, Marcelino07}. Figure \ref{fig:ColdCloud_Corino_Comparison} shows our carbon chain observations compared to the abundances found in TMC-1. We note that while the majority of our sources are in Perseus, which is ${\sim}55$\% further away than TMC-1, we have a beam size that is ${\sim}60$\% smaller than the smallest beams used in the original observations by \citet{Kaifu04} and therefore the two sets of observations probe similar physical scales.

We find that all our sources are underabundant in all species, often by several orders of magnitude, relative to the TMC-1 abundances. HC$_5$N is the most depleted (${\sim}1\times10^{4}$), followed by CCCS (${\sim}8\times10^{3}$) and l-C$_3$H (${\sim}4\times10^{3}$), while smaller decreases were seen for HC$_3$N (${\sim}3\times10^{3}$), CCS (${\sim}2\times10^{3}$), and C$_4$H (${\sim}2\times10^{2}$).

\begin{figure*}
\centering
\includegraphics[width=\linewidth]{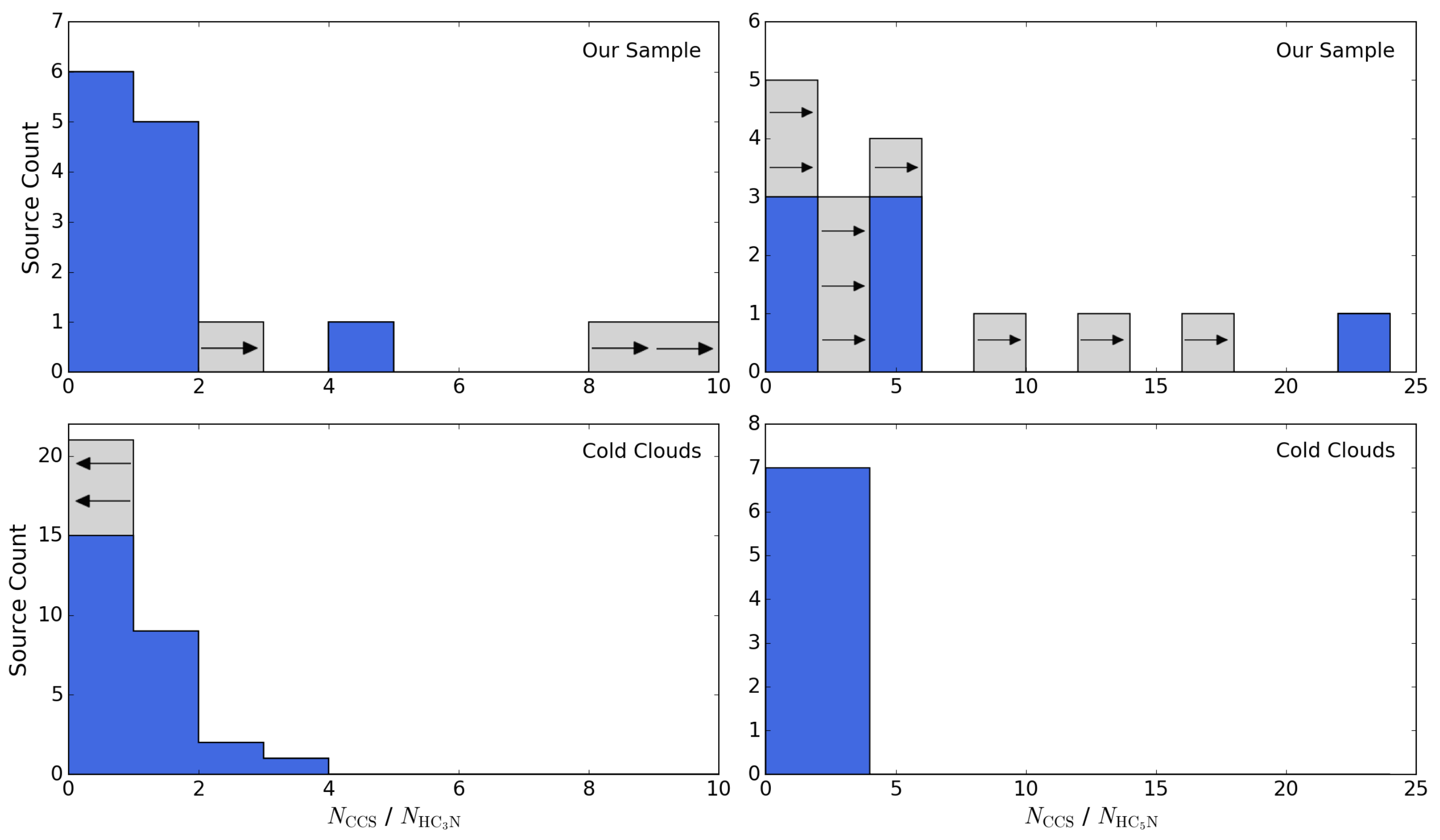}
\caption{Histograms of column density ratios of CCS versus HC$_3$N and HC$_5$N compared to ratios found in cold clouds by \citet{Suzuki92}. The top panels show the ratios in our protostellar sample and the bottom panels show the cold cloud ratios. In all panels, detections are shown in blue and non-detections in light gray with arrows indicating lower or upper limits.}
\label{fig:CCS_N_Bearing_Ratios}
\end{figure*}

The different depletion factors suggest a differentiation in carbon chain composition in the protostellar phase. Figure \ref{fig:CCS_N_Bearing_Ratios} directly compares two column density ratios, CCS/HC$_3$N and CCS/HC$_5$N, in cold clouds and protostellar envelopes using a different set of cloud observations. The cloud data include all dark cloud core sources in \citet{Suzuki92} for which ratios or ratio limits of CCS, HC$_3$N, and HC$_5$N could be derived. Figure \ref{fig:CCS_N_Bearing_Ratios} shows that while there is overlap between the cloud and protostellar samples, a large portion of the protostellar sources are overabundant in CCS or underabundant in HC$_3$N and HC$_5$N compared to cloud sources. These two samples are not observed in a homogeneous way and it is therefore difficult to quantify this difference, but it warrants further study to determine whether it is a general effect. An additional caveat is related to the fact that cyanpolyyne formation chemistry tends to proceed more slowly, causing nitrogen-bearing chains to be less abundant in the earlier phases of cold clouds \citep[e.g.,][]{Bergin97b}, while sulfur-bearing chains are more quickly formed \citep[e.g.,][]{Bergin97a, Aikawa01, Aikawa03, Aikawa05}. Thus, a cold cloud survey such as the one conducted by \citet{Suzuki92} could either be biased toward higher ratios of CCS/HC$_{\rm{n}}$N column density (since they were likely observing these systems when cyanpolyynes were still not yet all formed) or toward lower ratios if CCS has already begun freezing out.

\subsection{Comparison with WCCC Sources and Models}

\begin{figure*}[!htp]
\centering
\includegraphics[width=\linewidth]{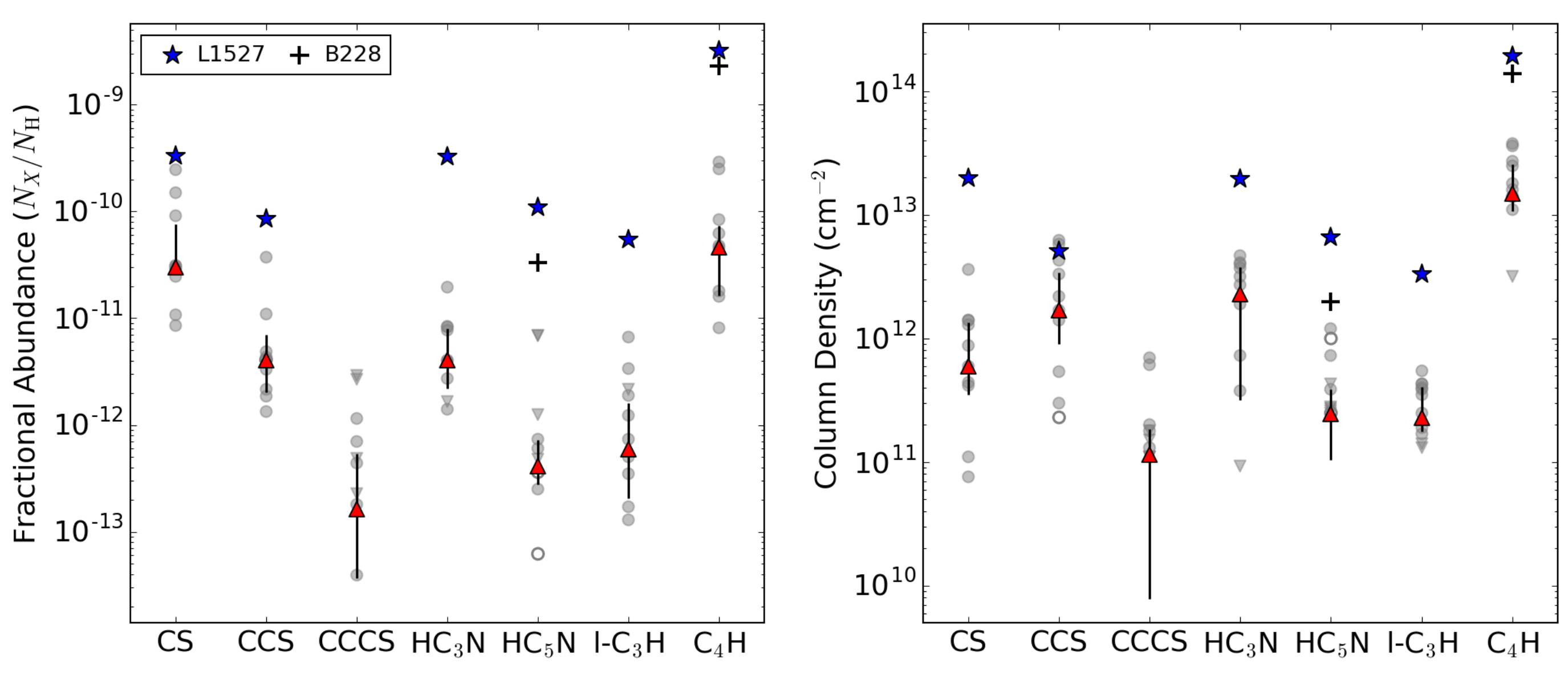}
\caption{Observed fractional abundances and column densities compared with observations of two WCCC sources. \textit{Left}: Median fractional abundances are represented by upward red triangles and error bars span the first and third quartiles derived by survival analysis. Individual sources are represented with gray circles for detections and downward triangles for upper limits, while open circle indicate tentative detections. \textit{Right}: Same as the left panel but with column densities instead of abundances.}
\label{fig:LukeWarm_Corino_Comparison}
\end{figure*}

We next compare our beam-averaged carbon chain abundances with observed abundances in WCCC sources to assess whether any of our sources show signs of WCCC chemistry. The left panel of Figure \ref{fig:LukeWarm_Corino_Comparison} shows a comparison of our observations with known WCCC sources taken from \citet{Hassel08, Hassel11}. The fractional abundances for B228 are based on observational data from the Nobeyama 45~m Telescope \citep{Sakai09B228}, which has a typical beam size of $20^{\prime \prime}$ at the observed frequencies. Since B228 is at a distance of 155~pc \citep{Lombardi08}, these observations are likely probing smaller spatial scales than our IRAM 30~m observations. The fractional abundances for L1527, which is at a distance of 137~pc \citep{Torres07}, are based on heterogeneous observations with beam sizes ranging from ${\sim}17^{\prime \prime}$--$37^{\prime \prime}$ \citep[see details within][]{Sakai08, Sakai09, Sakai09B228}. The data from species CCS and C$_4$H represent the same spatial scales as our IRAM 30~m observations, while those for HC$_3$N, HC$_5$N, and l-C$_3$H are likely probing somewhat smaller (${\sim}1.5$--$2.5\times$) spatial scales.

The medians for all molecules in our sample are underabundant relative to the WCCC observations by about two orders of magnitude, although there is significant scatter in carbon chain abundances among molecular families and individual protostars. L1455 IRS3 possesses the highest abundances for most molecules but no sources in our sample reach the elevated abundances observed in WCCCs, even when taking into account the factors of ${\sim}$2--5 arising from possible differences in beam dilution.

Since we have adopted a different approach for determining abundances than \citet{Hassel08, Hassel11}, it is possible that abundances have not been derived consistently between our sample and theirs. To mitigate this concern, the right panel of Figure \ref{fig:LukeWarm_Corino_Comparison} shows the same comparison but with column densities instead of fractional abundances. Our sources are still underabundant compared to WCCC sources, with the possible exception of CCS. We note that the abundance pattern is quite similar between our sample and the WCCC sources, but it is unclear whether this similarity is informative, since the pattern was also quite similar to what is observed in TMC-1.

\begin{figure}[!htp]
\centering
\includegraphics[width=\linewidth]{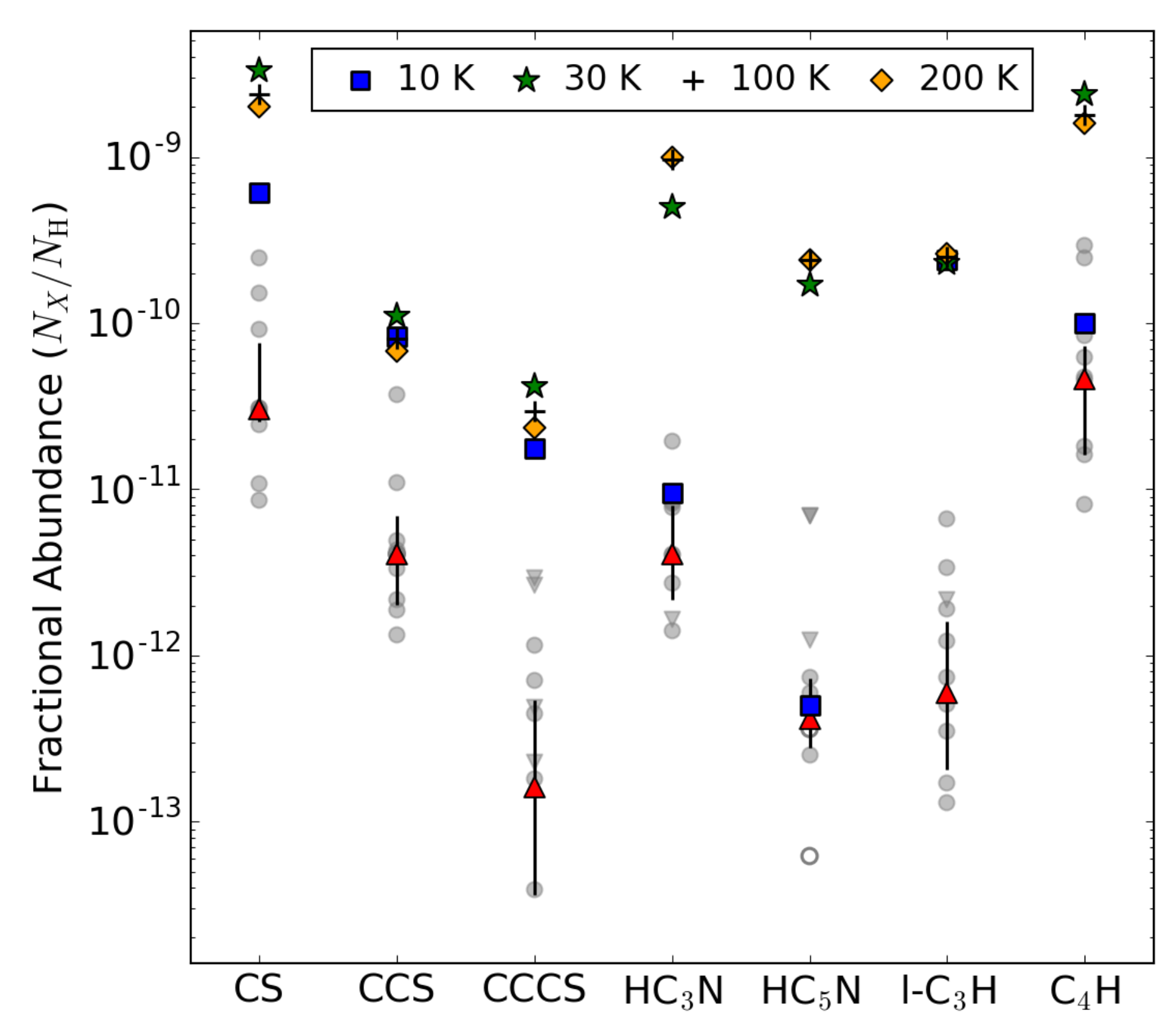}
\caption{Observed fractional abundances compared with various model predictions. Median fractional abundances are represented by upward red triangles and error bars span the first and third quartiles derived by survival analysis. Individual sources are represented with gray circles for detections and downward triangles for upper limits, while open circle indicate tentative detections.}
\label{fig:LukeWarm_Corino_Comparison_Models}
\end{figure}

Finally, we compare our sources with protostellar chemical models developed to explain the observed carbon chain abundances in WCCC sources. The protostellar model expectations shown in Figure \ref{fig:LukeWarm_Corino_Comparison_Models} are those taken from the latest and most comprehensive warm-up models of \citet{Hassel11}. The model fractional abundance for CCCS was provided by Hassel (private communication). The majority of our sources have abundances lower than the cold cloud model predictions ($T=10\,\rm{K}$) and no sources in our sample are consistent with a temperature of 30 K, which is that of WCCC sources, or the hot corino warm-up model expectations ($T=100, 200\,\rm{K}$). However, the abundance patterns look similar between the model and our sample. This could indicate that the observed carbon chains are formed \textit{in situ} but less efficiently than in the models, or alternatively, that these carbon chains were simply swept up as remnants from the cold cloud stages. To definitively distinguish between inheritance and WCCC would require spatially-resolved observations that can determine whether there are enhancements around the CH$_4$ desorption front or not.

\section{Conclusions}
\label{sec:conclusions}

Based on observations toward 16 young low-mass protostars using the IRAM 30 m telescope, we conclude the following:

\begin{enumerate}
\item Pure carbon chains, as well as nitrogen- and sulfur-bearing carbon chains, are found to be common at this stage (Class 0/I) of protostellar evolution. Specifically, we detect CCS, CCCS, HC$_3$N, HC$_5$N, l-C$_3$H, and C$_4$H in 88\%, 38\%, 75\%, 31\%, 81\%, and 88\% of the sources, respectively.
\item Carbon chain median abundance decreases with chain size for sulfur and nitrogen containing carbon chains and increases by two orders of magnitude from l-C$_3$H to C$_4$H. These findings are consistent with model expectations whether the carbon chains form through cold ion-molecule chemistry or warm carbon chain chemistry.
\item Carbon chain column densities span at least an order of magnitude, often two orders, across the sample. The distributions of carbon chains are quite broad, indicating the local conditions around protostellar hosts impact the carbon chain inventory.
\item We find significant correlations between molecules of the same carbon chain families as well as a strong correlation between the cyanpolyynes HC$_3$N and HC$_5$N and the pure hydrocarbon chains l-C$_3$H and C$_4$H. This latter correlation suggests that the production chemistry of C$_{\rm{n}}$H and the cyanpolyynes are closely related during low-mass star formation.
\item Carbon chain abundances for all sources in our sample are underabundant, often by several orders of magnitude, relative to those reported for cold cloud TMC~-1, WCCC sources, and warm-up model expectations.
\end{enumerate}

\acknowledgments

The authors thank the anonymous referee for valuable comments that improved the content and presentation of this work. This work has benefited from the helpful comments of David Wilner and Michael McCarthy. The study is based on observations with the IRAM 30~m telescope. IRAM is supported by INSU/CNRS (France), MPG (Germany), and IGN (Spain). K.I.{\"O}. acknowledges funding from an investigator award from the Simons Collaboration on the Origins of Life (SCOL). J.B.B. acknowledges funding from the National Science Foundation Graduate Research Fellowship under Grant DGE1144152.

%

\vspace{5mm}
\facilities{IRAM:30m}


\software{CLASS, \url{https://www.iram.fr/IRAMFR/GILDAS/}
          }



\pagebreak

\appendix

\section{Rotation Diagrams} \label{sec:rot_diagrams_appendix}

\begin{figure*}[!htp]
\centering
\includegraphics[scale=0.6]{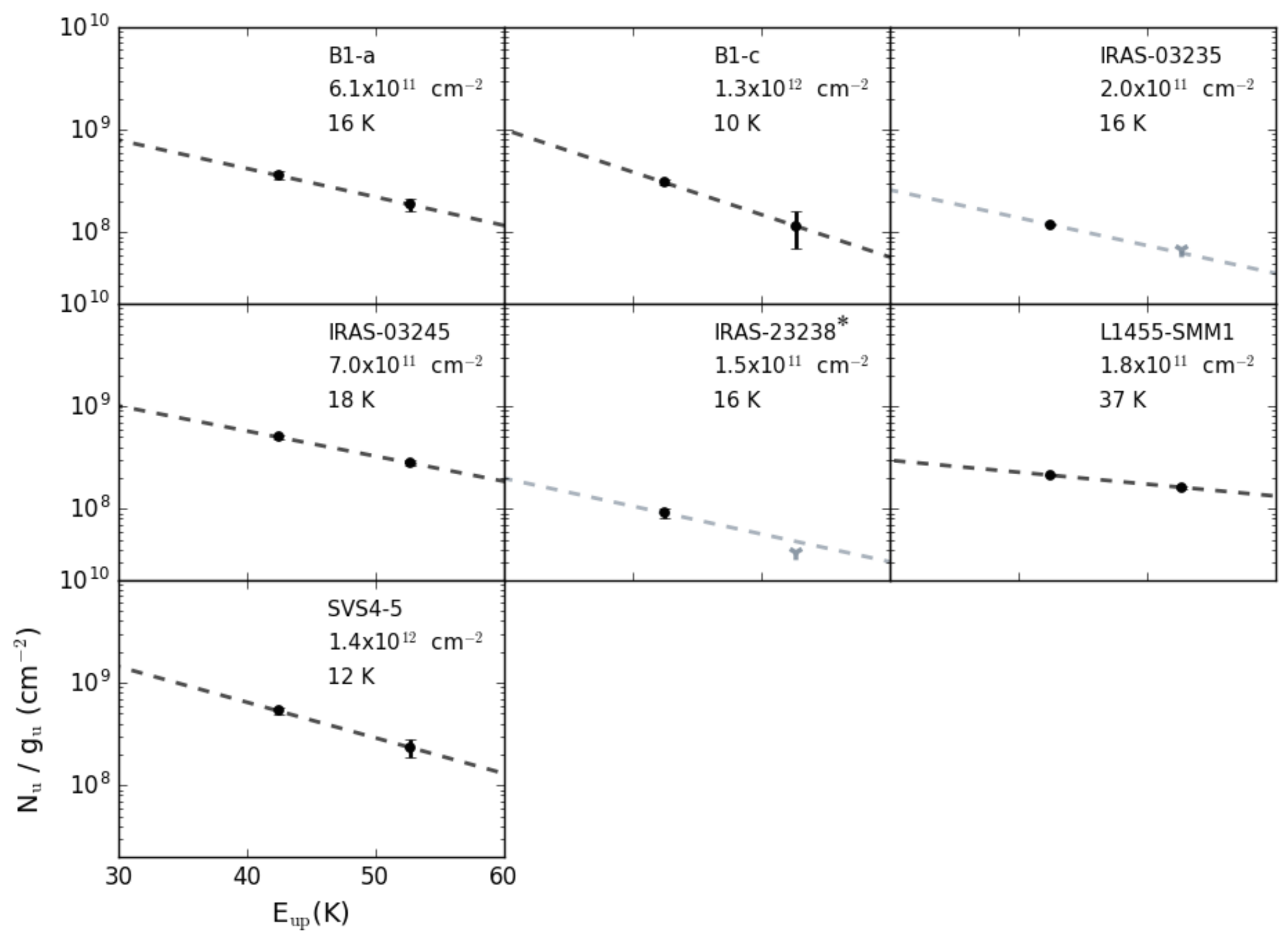}
\caption{Rotational diagrams for CCCS. Black circles represent detections and gray triangles indicate upper limits. Black dashed lines are fits to the data. When a line was unable to be fit, a rotational temperature was assumed as described in the text and is shown as a gray dashed line. Error bars are smaller than the symbol for most cases. A tentative CCCS detection in IRAS 23238 is denoted by an asterisk ($^*$).}
\label{fig:appendix_CCCS_RD}
\end{figure*}

\begin{figure*}[!htp]
\centering
\includegraphics[scale=0.6]{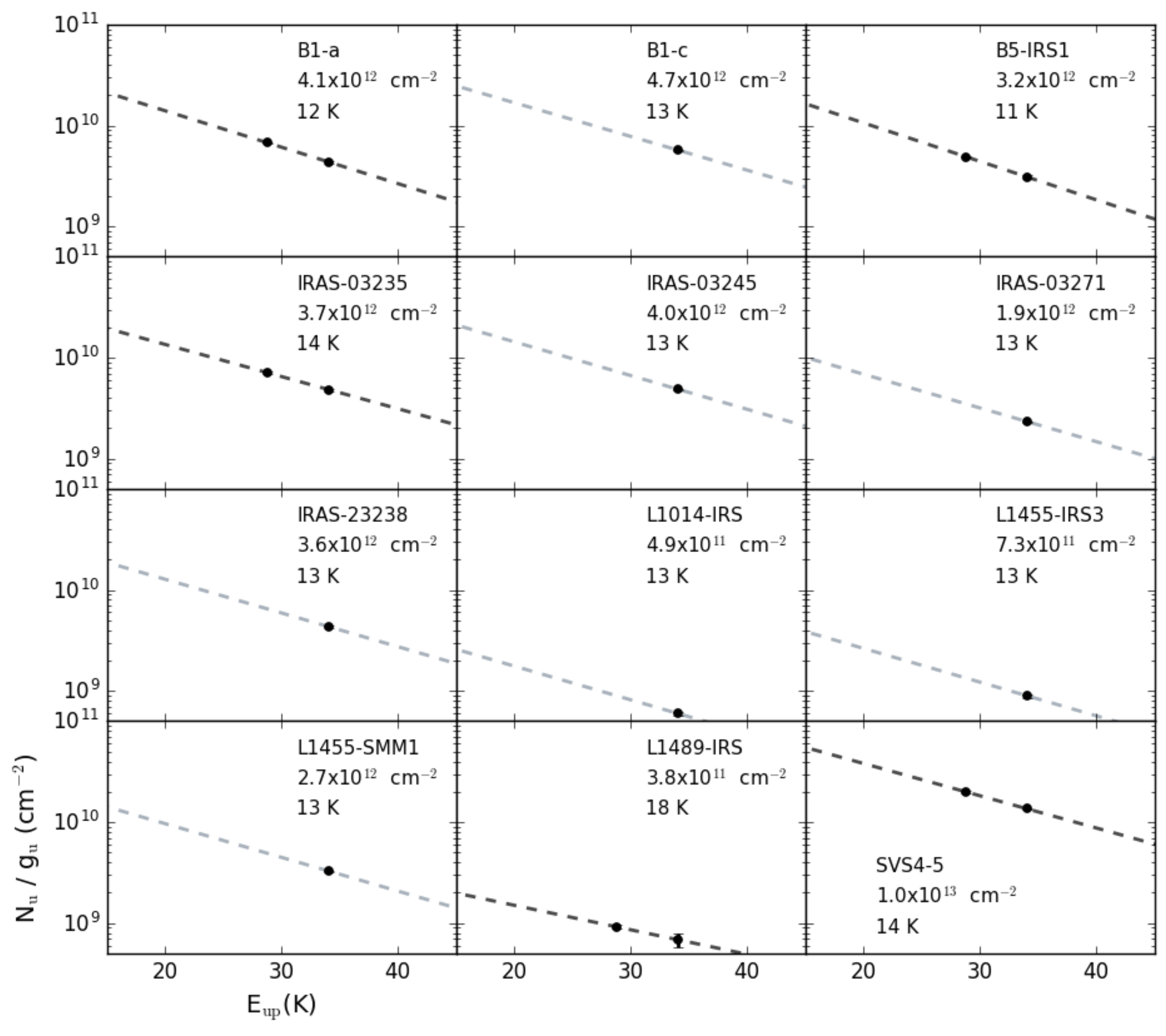}
\caption{Rotational diagrams for HC$_3$N. Black circles represent detections and gray triangles indicate upper limits. Black dashed lines are fits to the data. When a line was unable to be fit, a rotational temperature was assumed as described in the text and is shown as a gray dashed line. Error bars are smaller than the symbol for most cases.}
\label{fig:appendix_HC3N_RD}
\end{figure*}

\begin{figure*}[!htp]
\centering
\includegraphics[scale=0.65]{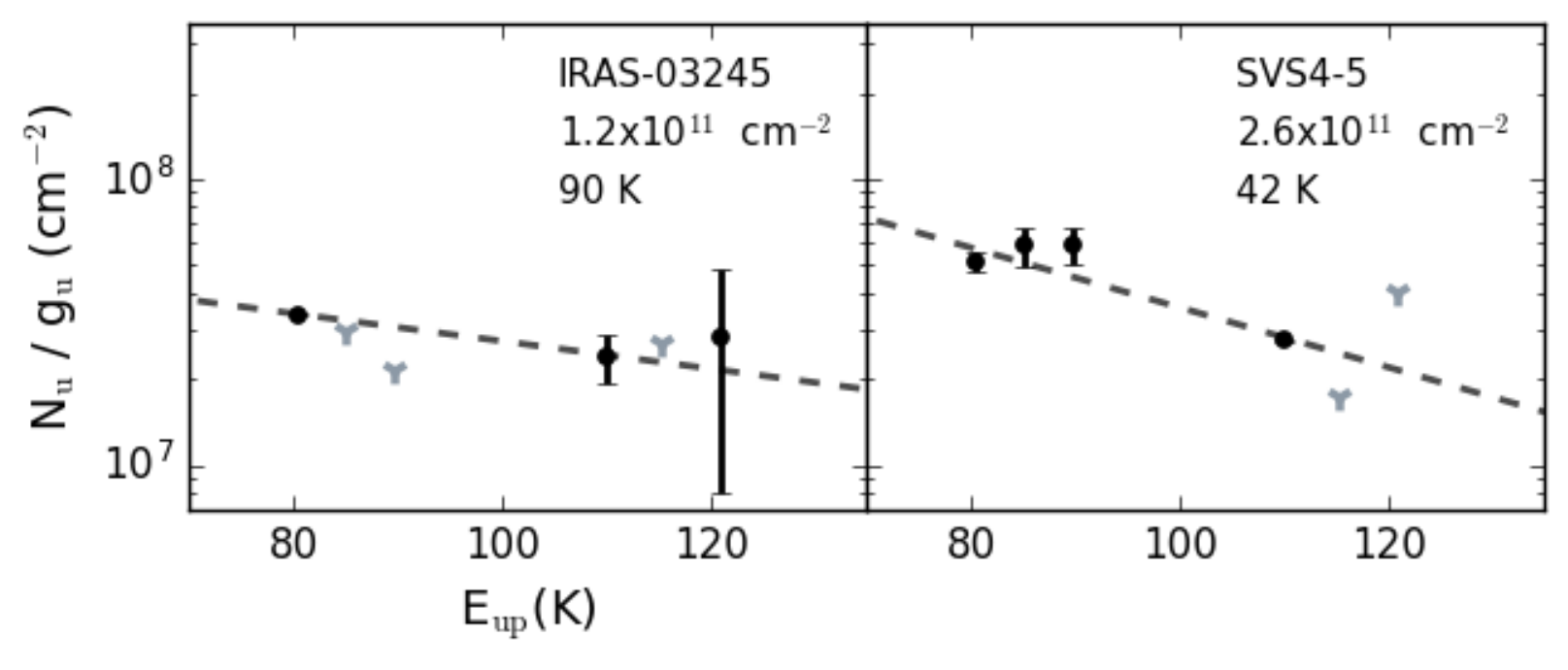}
\caption{Rotational diagrams for HC$_5$N. Black circles represent detections and gray triangles indicate upper limits. Black dashed lines are the fits to the data. When a line was unable to be fit, a rotational temperature was assumed as described in the text and is shown as a gray dashed line. Due to high fit uncertainties for IRAS 03245, the rotational temperature and column density shown here are treated as upper limits and we instead calculate column density using $T_{\rm{rot}} = 28$ K, as described in the text.}
\label{fig:appendix_HC5N_RD}
\end{figure*}

\section{Observed Transitions and Upper Limits} \label{sec:rotate}

\begin{deluxetable*}{lccc}[!htp]
\centering
\tablecaption{Integrated CS and Isotopologues C$^{34}$S, C$^{33}$S Intensities in K km s$^{-1}$\label{tab:CS_intensities}}
\tablehead{[-.3cm]
& C$^{34}$S / $96.413$ GHz & C$^{33}$S / $97.172$ GHz & CS / $97.981$ GHz  \\
Source & $J=2-1$ & $J=2-1$ & $J=2-1$}
\startdata 
L1448 IRS1\tablenotemark{$a$} & 0.028 [0.002] & 0.019 [0.002] & 0.535 [0.003] \\
IRAS 03235$+$3004 & 0.195 [0.005] & 0.021 [0.003] &1.151 [0.010] \\
IRAS 03245$+$3002 & 0.463 [0.004] & 0.060 [0.004] & 4.283 [0.008] \\
L1455 SMM1 & 0.436 [0.003] & 0.063 [0.003] & 2.951 [0.009] \\
L1455 IRS3 & 0.134 [0.003] & 0.025 [0.003] & 1.116 [0.003] \\
IRAS 03254$+$3050\tablenotemark{$a$} & 0.025 [0.003] & $<$0.010 & 0.454 [0.005] \\
IRAS 03271$+$3013 & 0.037 [0.003] & $<$0.007 & 0.758 [0.003] \\
B1-a & 1.158 [0.008] & 0.194 [0.008] & 6.355 [0.012] \\
B1-c\tablenotemark{$a$} & 0.406 [0.004] & 0.074 [0.005] & 2.572 [0.032] \\
B5 IRS 1 & 0.285 [0.008] & 0.041 [0.006] & 2.199 [0.008] \\
L1489 IRS & 0.082 [0.006] & 0.016 [0.004] & 1.526 [0.003] \\
IRAS 04108$+$2803 & 0.132 [0.004] & 0.023 [0.003] & 0.658 [0.003] \\
HH 300 & 0.143 [0.007] & 0.029 [0.006] & 0.745 [0.013] \\
SVS 4-5 & 1.579 [0.007] & 0.188 [0.006] & 13.62 [0.032] \\
L1014 IRS & 0.093 [0.002] & 0.012 [0.002] & 0.729 [0.002] \\
IRAS 23238$+$7401 & 0.241 [0.002] & 0.045 [0.003] & 2.407 [0.005] \\
\enddata
\tablenotetext{a}{B1-c shows weak self-absorption to the right side of lines, IRAS 03254$+$3050 shows significant self-absorption to the left of lines, and L1448 IRS1 shows weak self-absorption to the right of lines. \\}
\tablecomments{Uncertainties at the $1\sigma$ level are reported in brackets.}
\end{deluxetable*}

\begin{deluxetable*}{lccc}[!htp]
\centering
\tablecaption{Integrated CCS Intensities in K km s$^{-1}$\label{tab:CCS_intensities}}
\tablehead{[-.3cm]
& $93.870$ GHz & $99.867$ GHz & $113.410$ GHz  \\
Source & $J_N=8_7-7_6$ & $J_N=7_8-6_7$ & $J_N=8_9-7_8$}
\startdata 
L1448 IRS1 & $<$0.014 & $<$0.008 & $<$0.017 \\
IRAS 03235$+$3004 & 0.174 [0.005] & 0.053 [0.003] & 0.050 [0.007] \\
IRAS 03245$+$3002 & 0.309 [0.005] & 0.132 [0.003] &  0.121 [0.005]\\
L1455 SMM1 & 0.338 [0.004] & 0.114 [0.003] & 0.095 [0.006] \\
L1455 IRS3 & 0.103 [0.004] & 0.036 [0.003] & $<$0.022 \\
IRAS 03254$+$3050\tablenotemark{$a$} & 0.017 [0.004] & $<$0.012 & $<$0.025 \\
IRAS 03271$+$3013 & 0.023 [0.004] & $<$0.007 & $<$0.018  \\
B1-a & 0.476 [0.005] & 0.193 [0.004] & 0.142 [0.007] \\
B1-c & 0.321 [0.005] & 0.084 [0.003] & 0.056 [0.010] \\
B5 IRS 1 & 0.145 [0.008] & 0.068 [0.005] & 0.050 [0.012]  \\
L1489 IRS & 0.029 [0.007] & 0.019 [0.004] & $<$0.027 \\
IRAS 04108$+$2803 & 0.046 [0.004] & $<$0.010 & $<$0.020  \\
HH 300 & 0.066 [0.006] & 0.018 [0.003] & 0.013 [0.009] \\
SVS 4-5 & 0.312 [0.005] & 0.111 [0.002] & 0.106 [0.005]  \\
L1014 IRS & 0.076 [0.003] & 0.017[0.003] & $<$0.014 \\
IRAS 23238$+$7401 &  0.103 [0.002] & 0.032 [0.002] & $<$0.023 \\
\enddata
\tablenotetext{a}{Tentative single-line detection at $\gtrsim 4\sigma$. \\}
\tablecomments{Uncertainties at the $1\sigma$ level are reported in brackets.}
\end{deluxetable*}

\begin{deluxetable*}{lcc}[!htp]
\centering
\tablecaption{Integrated CCCS Intensities in K km s$^{-1}$\label{tab:CCCS_intensities}}
\tablehead{[-.3cm]
& $98.269$ GHz & $109.828$ GHz   \\
Source & $J = 17-16$ & $J = 19-18$} 
\startdata
L1448 IRS1 & $<$0.008 & $<$0.012\\
IRAS 03235$+$3004 & 0.015 [0.003] & $<$0.011 \\
IRAS 03245$+$3002 & 0.065 [0.003] & 0.045 [0.003] \\
L1455 SMM1 & 0.028 [0.003] & 0.026 [0.003] \\
L1455 IRS3 & $<$0.009 & $<$0.014  \\
IRAS 03254$+$3050 & $<$0.012 & $<$0.011 \\
IRAS 03271$+$3013 & $<$0.009 & $<$0.010\\
B1-a & 0.046 [0.002] & 0.030 [0.005] \\
B1-c & 0.040 [0.004] & 0.019 [0.003] \\
B5 IRS 1 & $<$0.014 & $<$0.023  \\
L1489 IRS & $<$0.009 & $<$0.014 \\
IRAS 04108$+$2803 & $<$0.012 & $<$0.010 \\
HH 300 & $<$0.013  & $<$0.022 \\
SVS 4-5 & 0.069 [0.003] & 0.038 [0.003]  \\
L1014 IRS & $<$0.008 & $<$0.009 \\
IRAS 23238$+$7401\tablenotemark{$a$} & 0.012 [0.003] & $<$0.006\\
 \enddata
 \tablenotetext{a}{Tentative single-line detection at $4\sigma$. \\}
 \tablecomments{Uncertainties at the $1\sigma$ level are reported in brackets.}
\end{deluxetable*}

\begin{deluxetable*}{lcccc}[!htp]
\centering
\tablecaption{Integrated l-C$_3$H Intensities in K km s$^{-1}$\label{tab:C3H_intensities}}
\tablehead{[-.3cm]
& $97.995$ GHz & $97.996$ GHz & $98.012$ GHz & $98.013$ GHz  \\
Source & $F = 5 - 4, l=e$ & $F = 4 - 3, l=e$ & $F= 5 - 4, l=f$ & $F = 4 -3, l=f$}
\startdata
L1448 IRS1 & $<$0.009 & $<$0.009 & $<$0.007 & $<$0.007 \\
IRAS 03235$+$3004 & 0.049 [0.004] & 0.034 [0.002] & 0.037 [0.004] & 0.046 [0.003] \\
IRAS 03245$+$3002 & 0.007 [0.002] & 0.009 [0.003] & 0.026 [0.003] & 0.012 [0.002] \\
L1455 SMM1 & 0.034 [0.004] & 0.016 [0.003] & 0.038 [0.004] & 0.020 [0.003]  \\
L1455 IRS3 & 0.017 [0.003] & 0.016 [0.002] & 0.022 [0.003] & 0.021 [0.003]  \\
IRAS 03254$+$3050 & $<$0.012 & $<$0.012 &  0.016 [0.003] & $<$0.012 \\
IRAS 03271$+$3013 & 0.037 [0.003] & $<$0.012 & $<$0.008 & $<$0.009\\
B1-a & 0.032 [0.003] & 0.028 [0.003] & 0.025 [0.003] & 0.033 [0.003]\\
B1-c & 0.033 [0.003] & 0.033 [0.004] & 0.026 [0.004] & 0.036 [0.003]\\
B5 IRS 1 & 0.041 [0.005] & 0.030 [0.005] & 0.031 [0.006] & 0.020 [0.006]  \\
L1489 IRS & $<$0.011 & $<$0.011 & $<$0.013 & $<$0.013 \\
IRAS 04108$+$2803 & $<$0.009 & $<$0.009 &  0.017 [0.003] & 0.011 [0.005] \\
HH 300\tablenotemark{$a$} & 0.010 [0.004] & $<$0.014 &  0.012 [0.004] & $<$0.012\\
SVS 4-5 & 0.036 [0.003] & 0.035 [0.005] & 0.027 [0.002] & 0.027 [0.003]  \\
L1014 IRS & 0.010 [0.002] & 0.028 [0.003] & 0.011 [0.001] & 0.009 [0.003] \\
IRAS 23238$+$7401 & 0.021 [0.002] & 0.017 [0.004] & 0.020 [0.002] & 0.008 [0.002]  \\
 \enddata 
 \tablecomments{Uncertainties at the $1\sigma$ level are reported in brackets. All transitions are $J = 9/2 - 7/2$. }
 \tablenotetext{a}{Source does not meet our criteria for a detection.}
\end{deluxetable*}

\begin{deluxetable*}{lccc}[!htp]
\centering
\tablecaption{Integrated HC$_3$N Intensities in K km s$^{-1}$\label{tab:HC3N_intensities}}
\tablehead{[-.3cm]
& $100.076$ GHz & $109.174$ GHz  \\
Source & $J=11-10$ & $J=12-11$}
\startdata
L1448 IRS1 & \ldots & $<$0.009  \\
IRAS 03235$+$3004 & 0.649 [0.003] & 0.526 [0.003] \\
IRAS 03245$+$3002 & \ldots & 0.531 [0.002]  \\
L1455 SMM1 & \ldots &  0.357 [0.002] \\
L1455 IRS3 & \ldots & 0.097 [0.004] \\
IRAS 03254$+$3050 & \ldots & $<$0.012 \\
IRAS 03271$+$3013 & \ldots & 0.253 [0.004] \\
B1-a & 0.615 [0.004] & 0.474 [0.003] \\
B1-c & \ldots & 0.621 [0.005] \\
B5 IRS 1 & 0.448 [0.005] & 0.337 [0.007]  \\
L1489 IRS & 0.084 [0.005] & 0.074 [0.006] \\
IRAS 04108$+$2803 & $<$0.012 & $<$0.008  \\
HH 300 & \ldots & $<$0.018 \\
SVS 4-5 & 1.823 [0.007] & 1.475 [0.008]    \\
L1014 IRS & \ldots & 0.065 [0.003] \\
IRAS 23238$+$7401 & \ldots & 0.471 [0.003] \\
\enddata
\tablecomments{Uncertainties at the $1\sigma$ level are reported in brackets. Dots (\ldots) indicate that the 100.076 GHz transition was not contained in the spectral window for that particular source.}
\end{deluxetable*}

\begin{deluxetable*}{lccccccc}[!htp]
\centering
\tablecaption{Integrated HC$_5$N Intensities in K km s$^{-1}$\label{tab:HC5N_intensities}}
\tablehead{[-.3cm]
& $93.188$ GHz & $95.850$ GHz & $98.513$ GHZ & $109.161$ GHz & $111.823$ GHz & $114.485$ GHz  \\
Source & $J=35-34$ & $J=36-35$ & $J=37-36$ & $J=41-40$ & $J=42-41$ & $J=43-42$}
\startdata 
L1448 IRS1 & $<$0.009 & $<$0.012 & $<$0.010 & $<$0.009 & $<$0.014 & $<$0.024 \\
IRAS 03235$+$3004 & $<$0.017 & $<$0.017 & $<$0.010 & $<$0.012 & $<$0.012 & $<$0.023 \\
IRAS 03245$+$3002 & 0.011 [0.005] & $<$0.011 & $<$0.009 & 0.012 [0.004] & $<$0.014 & 0.012 [0.003]  \\
L1455 SMM1 & $<$0.014 & $<$0.014 & $<$0.010 & $<$0.012 & $<$0.013 & $<$0.022 \\
L1455 IRS3 & $<$0.012 & $<$0.014 & $<$0.010 & $<$0.011 & $<$0.012 & $<$0.024 \\
IRAS 03254$+$3050 & $<$0.015 & $<$0.014 & 0.015 [0.003] & $<$0.009 & $<$0.016 & $<$0.025 \\
IRAS 03271$+$3013 & $<$0.014 & $<$0.012 & $<$0.010 & $<$0.011 & $<$0.016 & $<$0.031 \\
B1-a\tablenotemark{$a$}  &  $<$0.012 & $<$0.015 & 0.009 [0.003] & $<$0.010 & $<$0.013 & $<$0.030  \\
B1-c & 0.017 [0.004] & $<$0.014 & $<$0.009 & $<$0.015  & $<$0.019 & 0.033 [0.011]\\
B5 IRS 1\tablenotemark{$a$}  & $<$0.031 & $<$0.023 & 0.012 [0.005] & 0.024 [0.006] & $<$0.027 & $<$0.052 \\
L1489 IRS & $<$0.020 & $<$0.022 & $<$0.016 & $<$0.015 & $<$0.021 & $<$0.029 \\
IRAS 04108$+$2803 & $<$0.016 & $<$0.014 & $<$0.009 & $<$0.013 & $<$0.012 & $<$0.028 \\
HH 300 & $<$0.016 & $<$0.015 & $<$0.013 & $<$0.018 & $<$0.020 & $<$0.026 \\
SVS 4-5 & 0.019 [0.005] & 0.023 [0.005] &  0.024 [0.002] & 0.014 [0.004]  & $<$0.009 & $<$0.022  \\
L1014 IRS & $<$0.011 & $<$0.010 & $<$0.006 & $<$0.009 & $<$0.012 & $<$0.022 \\
IRAS 23238$+$7401 & $<$0.012 & 0.016 [0.003] & 0.014 [0.002] & $<$0.010 & $<$0.009 & $<$0.021\\
\enddata
\tablenotetext{a}{Tentative single-line detections at $3\sigma$ and $4\sigma$ for B1-a and B5 IRS1, respectively. \\}
\tablecomments{Uncertainties at the $1\sigma$ level are reported in brackets.}
\end{deluxetable*}

\newpage


\begin{deluxetable*}{lccccccccc}
\tablecaption{FWHMs of Detected Lines in km s$^{-1}$\label{tab:FWHMs_Detected_Lines_1}}
\tablehead{[-.3cm]
 Source  & $\nu$ (GHz) &  L1448 IRS1 & IRAS 03235 &  IRAS 03245 &  L1455 SMM1 &  L1455 IRS3 &  IRAS 03254 & IRAS 03271 & B1-a \\[-.6cm]}
\startdata
CS              & $97.981$ & 1.03 & 1.16 & 1.61 & 1.39 & 1.02 & 0.98 & 1.92 & 2.19  \\
C$^{34}$S &  $96.413$ & 1.00 & 0.75 & 1.43 & 1.32 & 0.78 & 0.61 & 2.39 & 1.42 \\
C$^{33}$S &  $97.172$ & 3.25 & 0.75 & 1.39 & 1.59 & 1.15 & \ldots & \ldots & 1.55  \\
\hline \\[-.6cm]
CCS &  $93.870$          &  \ldots & 0.62  & 1.13 &  1.12  & 0.62 & 0.62 & 1.15 & 1.06\\
       &  $99.867$         &  \ldots & 0.59  & 1.29 &  0.95   & 1.15 & \ldots & \ldots & 1.39\\
       &   $113.410$      &  \ldots & 0.79  & 1.01 & 0.90 & \ldots & \ldots & \ldots & 1.12\\
\hline \\[-.6cm]
CCCS & $98.269$        & \ldots & 0.60  & 1.57 & 0.60  & \ldots & \ldots  & \ldots & 1.43 \\
         &  $109.828$    & \ldots & \ldots & 1.21 & 1.02 & \ldots &  \ldots & \ldots & 1.40 \\
\hline \\[-.6cm]
l-C$_3$H &  $97.995$ & \ldots & 1.14 & 0.60 & 1.43 & 1.01 & \ldots & 3.51 & 1.26 \\
               & $97.996$  & \ldots & 1.05 & 0.60 & 1.10 & 0.90 & \ldots & \ldots & 1.51 \\
               & $98.012$  & \ldots & 1.02 & 1.30 & 2.00 & 1.82 & \ldots & \ldots & 1.09 \\
               &  $98.013$ & \ldots & 1.64 & 1.07 & 1.48 & 1.12 & \ldots & \ldots & 1.67 \\
\hline \\[-.6cm]
HC$_3$N  & $100.076$ & \ldots & 1.11 & \ldots & \ldots & \ldots & \ldots & \ldots & 1.27 \\
                & $109.174$ & \ldots & 1.13 & 1.33   & 0.96   &  0.95  & \ldots & 1.04   &  1.29 \\
\hline \\[-.6cm]
HC$_5$N & $93.188$   & \ldots & \ldots   & \ldots & \ldots & \ldots & \ldots & \ldots & \ldots \\
               & $95.850$   & \ldots & \ldots   & \ldots &  \ldots & \ldots & \ldots & \ldots & \ldots \\
               & $98.513$   & \ldots & \ldots   & \ldots & \ldots   & \ldots & 1.00   & \ldots & 1.39 \\
               & $109.161$ & \ldots & \ldots  & 1.20   & \ldots & \ldots   & \ldots  & \ldots & \ldots \\
               & $111.823$ & \ldots & \ldots & \ldots & \ldots & \ldots   & \ldots & \ldots & \ldots \\
               & $114.485$ & \ldots & \ldots & 3.20   & \ldots & \ldots & \ldots & \ldots & \ldots \\
\enddata
\end{deluxetable*}

\addtocounter{table}{-1}
\begin{deluxetable*}{lccccccccc}
\tablecaption{FWHMs of Detected Lines in km s$^{-1}$ \textit{(continued)}\label{tab:FWHMs_Detected_Lines_2}}
\tablehead{[-.3cm]
Source  & $\nu$ (GHz)  &  B1-c &  B5 IRS 1 & L1489 IRS & IRAS 04108 &  HH 300 &  SVS 4-5 & L1014 IRS & IRAS 23238 \\[-.6cm]}
\startdata
CS              & $97.981$ & 1.49 & 1.10 & 1.68 & 1.52  & 1.23 & 3.31   & 1.19   & 1.90 \\
C$^{34}$S & $96.413$ & 1.28 & 0.71 & 1.77 & 0.90  & 0.95 & 2.61   & 0.91   & 1.21 \\
C$^{33}$S & $97.172$ & 1.10 & 0.61 & 1.37 & 1.68  & 1.06 & 2.53 & 1.38 & 1.63 \\
\hline \\[-.6cm]
CCS &  $93.870$           & 0.67   &  0.70   &   0.62 & 0.80 & 0.65 & 2.01   & 0.62   & 0.81 \\
       &  $99.867$           & 1.03   & 0.94  &  1.76  &\ldots & 1.04 & 2.05 & 1.06   & 1.14  \\
       &   $113.410$        & 0.85   & 1.07  & \ldots &\ldots & 0.92 & 1.90 & \ldots & \ldots \\
\hline \\[-.6cm]
CCCS & $98.269$         & 1.22 & \ldots  &  \ldots & \ldots & \ldots & 3.39 & \ldots & 1.07 \\
         & $109.828$       & 0.53 &  \ldots & \ldots & \ldots & \ldots & 2.49 & \ldots & \ldots \\
\hline \\[-.6cm]
l-C$_3$H &  $97.995$ & 1.21 & 1.30 &  \ldots & \ldots & 0.76   & 1.84   & 0.94 & 1.15 \\
               & $97.996$  & 1.32 & 2.34 &  \ldots & \ldots & \ldots & 2.67 & 1.60 & 0.95 \\
               & $98.012$  & 1.07 & 0.99   &  \ldots & 1.58 & 2.02   & 2.13 & 0.60 & 0.90 \\
               &  $98.013$ & 1.23 & 1.28   &  \ldots & 1.30 & \ldots & 2.13 & 1.07 & 0.79 \\
\hline \\[-.6cm]
HC$_3$N & $100.076$ & \ldots & 1.03 &  1.23 & \ldots & \ldots & 2.49 & \ldots & \ldots \\
               & $109.174$ & 1.24   & 1.10 &  1.25 & \ldots & \ldots & 2.54  & 0.94 & 1.09 \\
\hline \\[-.6cm]
HC$_5$N & $93.188$   & 1.00 & \ldots   & \ldots & \ldots & \ldots & 0.91 & \ldots & \ldots \\
               & $95.850$   & \ldots & \ldots  & \ldots & \ldots & \ldots & 1.86 & \ldots & 1.47 \\
               & $98.513$   & \ldots & 0.83   & \ldots & \ldots & \ldots & 2.51 & \ldots   & 1.90 \\
               & $109.161$ & \ldots & 1.44    & \ldots & \ldots & \ldots & 0.90 & \ldots & \ldots \\
               & $111.823$ & \ldots & \ldots  & \ldots & \ldots & \ldots & \ldots & \ldots & \ldots \\
               & $114.485$ & 1.24 & \ldots    & \ldots & \ldots & \ldots & \ldots & \ldots & \ldots \\
\enddata
\end{deluxetable*}




\bibliography{Thesis_References}



\end{document}